\documentclass[prc, amsfonts, amssymb, amsmath, preprintnumbers, twocolumn, showkeys, nofootinbib,superscriptaddress,floatfix]{revtex4-1}%

\usepackage[english]{babel}
\usepackage[colorinlistoftodos, color=green!40, prependcaption]{todonotes}
\usepackage{amsmath}
\usepackage{amssymb}
\usepackage{booktabs} %
\usepackage{hyperref}

\usepackage[compat=1.1.0]{tikz-feynman}
\usepackage{contour}
\usepackage{caption}
\usepackage{subcaption}

\usepackage{amsthm}
\usepackage{mathtools}
\usepackage{physics}
\usepackage{xcolor}
\usepackage{graphicx}
\usepackage[left=23mm,right=13mm,top=35mm,columnsep=15pt]{geometry} 
\usepackage{adjustbox}
\usepackage{placeins}
\usepackage[T1]{fontenc}
\usepackage{lipsum}
\usepackage{csquotes}
\usepackage{listings}

\usepackage{csquotes}

\graphicspath{{Images/}}

\usepackage{xspace}

\usepackage{symbols}

\begin{document} 

\title{\boldmath Higgs Production and $\Hcc$ at a Future Muon-Ion Collider}

\author{Pavan Ahluwalia}
\affiliation{Department of Physics, Southern Methodist University, Dallas, TX 75275, USA}
\affiliation{Harmony School of Innovation, Fort Worth, TX 76123, USA}
\author{Justine Choi}
\affiliation{Department of Physics, Southern Methodist University, Dallas, TX 75275, USA}
\affiliation{Highland Park High School, Highland Park, TX 75205, USA}
\author{Simba Masters}
\affiliation{Department of Physics, Southern Methodist University, Dallas, TX 75275, USA}
\author{Austin Mullins}
\affiliation{Department of Physics, Southern Methodist University, Dallas, TX 75275, USA}
\author{Noah Randall}
\affiliation{Department of Physics, Southern Methodist University, Dallas, TX 75275, USA}
\author{Stephen J. Sekula}
\email{corresponding author: stephen.sekula@snolab.ca}
\affiliation{Department of Physics, Southern Methodist University, Dallas, TX 75275, USA}
\affiliation{SNOLAB, Lively, ON P3Y 1N2, Canada}
\affiliation{Department of Physics, Engineering Physics, and Astronomy, Queen's University, Kingston, ON K7L 3N6, Canada}

\begin{abstract}
We investigate the production of Higgs bosons in neutral- and charged-current interactions at a future Muon-Ion Collider (MuIC) with emphasis on $\Hcc$ decay. In particular, we focus on the leading production mechanisms and explore the impact of a range of Parton Distribution Function (PDF) scenarios, with emphasis on the role of heavy intrinsic quark flavor. We emphasize estimates of non-resonant di-charm background production and explore the impact of this background on a possible future measurement of $\Hcc$ at a future Muon-Ion Collider facility.
\end{abstract}

\preprint{SMU-HEP-22-11}

\maketitle

\section{Introduction}

The Electron-Ion Collider at Brookhaven National Laboratory~\cite{AbdulKhalek:2021gbh,AbdulKhalek:2022erw} will open an entirely new window on hadron structure through the use of high-intensity, high-energy electron and ion collisions. A recent proposal for later upgrading the EIC to a Muon-Ion Collider (MuIC) has been presented~\cite{Acosta:2021qpx} and a case is being built within the community for such an upgrade path beyond the nominal EIC program~\cite{Acosta:2022ejc}. A MuIC would not only provide a useful step forward toward the eventual idea of a dedicated muon collider, but itself would facilitate a novel physics program that complements the High-Luminosity Large Hadron Collider (HL-LHC) and other future accelerator complexes (c.f. Ref.~\cite{Shiltsev:2019rfl}). In particular, the center-of-mass energy possible in such a machine would facilitate a program that includes Higgs physics.

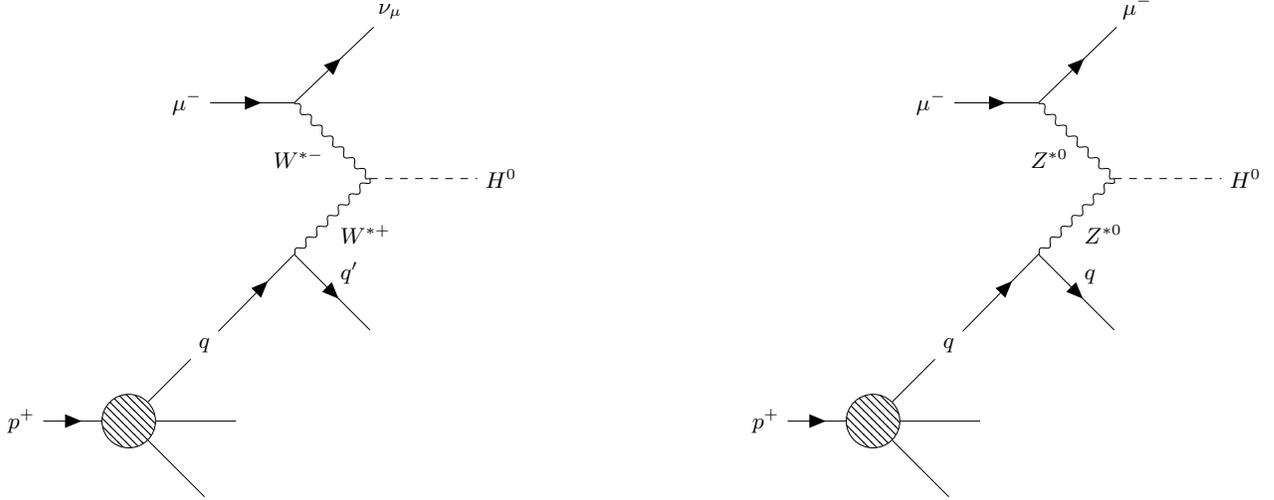
\begin{figure}[htbp]
    \begin{subfigure}[t]{0.45\textwidth}
      \scalebox{0.95}{
    \begin{tikzpicture}[scale=0.5]
     \begin{feynman}
      \vertex(a) {\(\mu^{-}\)};
      \vertex[right=of a] (V1);
      \vertex[above right=of V1] (f1){\(\nu_{\mu}\)};
      \vertex[below right=of V1] (H);
      \vertex[right=of H] (Hdecay){\(H^0\)};
      \vertex[below left=of H] (V2);
      \vertex[below left=of V2] (q){\(q\)};
      \vertex[below right=of V2] (charm);
      \vertex[blob, below left=of q] (dis) {\contour{gray}{}};
      \vertex[left=of dis] (p) {\(p^+\)};
      \vertex[right=of dis] (frag1);
      \vertex[below right=of dis] (frag2);
      \diagram* {(a)-- [fermion] (V1)-- [fermion] (f1),
      (V1)-- [boson,edge label'=\(W^{*-}\)] (H),
      (H)-- [scalar] (Hdecay),
      (V2)-- [boson,edge label'=\(W^{*+}\)] (H),
      (V2)-- [anti fermion] (q),
      (V2)-- [fermion, edge label=$q^{\prime}$] (charm),
      (p)-- [fermion] (dis) -- {(q), (frag1), (frag2)}
      };
     \end{feynman}
    \end{tikzpicture}
  }
  \end{subfigure}
\hfill
  \begin{subfigure}[t]{0.45\textwidth}
      \scalebox{0.95}{
    \begin{tikzpicture}[scale=0.5]
     \begin{feynman}
      \vertex(a) {\(\mu^{-}\)};
      \vertex[right=of a] (V1);
      \vertex[above right=of V1] (f1){\(\mu^-\)};
      \vertex[below right=of V1] (H);
      \vertex[right=of H] (Hdecay){\(H^0\)};
      \vertex[below left=of H] (V2);
      \vertex[below left=of V2] (q){\(q\)};
      \vertex[below right=of V2] (charm);
      \vertex[blob, below left=of q] (dis) {\contour{gray}{}};
      \vertex[left=of dis] (p) {\(p^+\)};
      \vertex[right=of dis] (frag1);
      \vertex[below right=of dis] (frag2);
      \diagram* {(a)-- [fermion] (V1)-- [fermion] (f1),
      (V1)-- [boson,edge label'=\(Z^{*0}\)] (H),
      (H)-- [scalar] (Hdecay),
      (V2)-- [boson,edge label'=\(Z^{*0}\)] (H),
      (V2)-- [anti fermion] (q),
      (V2)-- [fermion, edge label=$q$] (charm),
      (p)-- [fermion] (dis) -- {(q), (frag1), (frag2)}
      };
     \end{feynman}
    \end{tikzpicture}
  }
  \end{subfigure}

\caption{\label{fig:dis} Leading-order diagram for the production
of a single final-state Higgs particle in charged-current (left) and neutral-current (right) muon-proton DIS. In both cases a particle jet is expected both from the proton remnant and from the scattered quark that participated in the DIS interaction.
}
\end{figure}

In this note, we explore in additional detail the production of Higgs particles at such a collider. We focus specifically on $\Hcc$ and di-charm production that would serve as a background to the Higgs reconstruction. The decay \Hcc\ is a key part of the ongoing LHC program (c.f. Refs~\cite{ATLAS:2022ers,CMS:2022psv}) but poses a challenge in the LHC proton-collider environment. Mechanisms for production of the Higgs boson at a lepton-hadron collider are well-understood in the context of past such collider studies. A MuIC operated at, for example, a reasonable center-of-mass energy ($E_{\mu}=960~\GeV$, $E_{p}=275~\GeV$, $\sqrt{s}\approx 1000~\GeV$, achievable without a complete overhaul of the EIC complex) would allow for single-Higgs production dominated by charged-current (CC) and neutral-current (NC) deep-inelastic scattering (DIS) involving light-flavor quarks. This is illustrated in Fig.~\ref{fig:dis}.

The cross-sections for the above processes were estimated in Ref.~\cite{Acosta:2022ejc} for a $960\gev \times 275\gev$ muon-proton collider. Estimates were made at leading order (LO) in the matrix element and next-to-leading order (NLO) in the parton distribution function (PDF). The matrix element was computed using \textsc{MadGraph 3.3.1} and the \texttt{sm-no\_b\_mass} model (a "five-flavor scheme" in which the quarks contributing to the proton are assumed to all be massless except the top quark). The PDF set used was \texttt{PDF4LHC15\_nlo\_mc\_pdfas}. A number of kinematic selections are set implicitly by \textsc{MadGraph}; these are disabled (e.g. the minimum jet \pt\ is set to 0 instead of 20~GeV, etc.). For certain processes (c.f. Section~\ref{sec:backgrounds}) we do employ generator-level kinematic criteria. 

\begin{figure}[htbp]
      \scalebox{0.95}{
    \begin{tikzpicture}[scale=0.5]
     \begin{feynman}
      \vertex(a) {\(\mu^{-}\)};
      \vertex[right=of a] (V1);
      \vertex[above right=of V1] (f1){\(\nu_{\mu}\)};
      \vertex[below right=of V1] (H);
      \vertex[right=of H] (Hdecay){\(H^0\)};
      \vertex[below left=of H] (V2);
      \vertex[below left=of V2] (q){\(\bbar\)};
      \vertex[below right=of V2] (charm);
      \vertex[blob, below left=of q] (dis) {\contour{gray}{}};
      \vertex[left=of dis] (p) {\(p^+\)};
      \vertex[right=of dis] (frag1);
      \vertex[below right=of dis] (frag2);
      \diagram* {(a)-- [fermion] (V1)-- [fermion] (f1),
      (V1)-- [boson,edge label'=\(W^{*-}\)] (H),
      (H)-- [scalar] (Hdecay),
      (V2)-- [boson,edge label'=\(W^{*+}\)] (H),
      (V2)-- [anti fermion] (q),
      (V2)-- [fermion, edge label=$\overline{t}$] (charm),
      (p)-- [fermion] (dis) -- {(q), (frag1), (frag2)}
      };
     \end{feynman}
    \end{tikzpicture}
  }

\caption{\label{fig:tH} Leading-order diagram for the production
of a single final-state Higgs particle in top-Higgs associated production via charged-current muon-proton DIS. 
}
\end{figure}
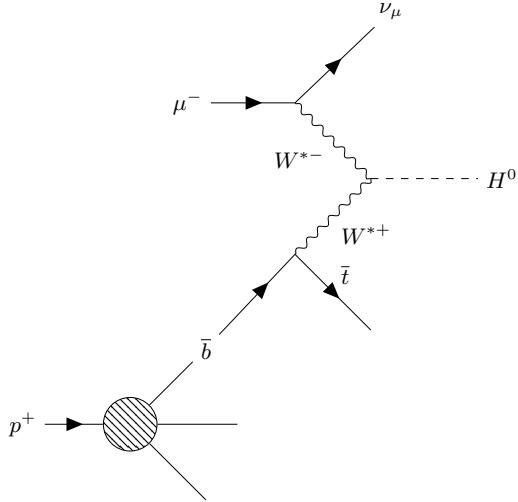

For zero beam polarization, the authors of Ref.~\cite{Acosta:2022ejc} determined the CC (NC) DIS cross-section to be $65~\mathrm{fb}$ ($12~\mathrm{fb}$). In addition, the authors did a dedicated simulation of a top-associated Higgs production channel that is strongly influenced by the bottom quark content in the proton and the modeling of the proton's gluon content (which can lead to $g^* \to b \bbar$ splitting). They estimated the cross-section for this process (Fig.~\ref{fig:tH}) to be $0.0158~\fb$ (also at zero polarization). 

The authors of the study also estimated the uncertainties associated with these production mechanisms owing to the scale and the combined PDF and $\alpha_s$ uncertainties. They estimated the scale uncertainty in the CC and NC DIS processes to be at the level of 6\% and the combined PDF and $\alpha_s$ uncertainty to be at the level of 1\%. For the top-Higgs associated production mechanism, these were estimated to be at the level of $1-2\%$ and $14\%$, respectively.

In this note, we reproduce some of the key results in Ref.~\cite{Acosta:2022ejc} and explore in more detail the effect of PDFs on various production mechanism (especially as regards the modeling of heavy flavor content in the proton) as well as a dedicated study of the $H\to\overline{c}$ final state. All simulations for this study were run using \textsc{MadGraph}~\cite{Alwall:2014hca, Frederix:2018nkq}, \textsc{Pythia 8.306}~\cite{Bierlich:2022pfr, Sjostrand:2006za}, and \textsc{Delphes 3}~\cite{deFavereau:2013fsa}.

To facilitate the simulation of lepton-hadron collisions, we implement a custom calculation of $Q^2$ that utilizes the initial- and final-state leptons to ascertain the momentum transfer-squared to the proton. This is documented in Appendix~\ref{sec:simulation_details} and is based on conversations with some of the authors of Ref.~\cite{Acosta:2022ejc}. In addition, to facilitate the hadronization stage we employ a newer model of color flow for the beam remnant that is implemented in \textsc{Pythia8}, also documented in the appendix.

\onecolumngrid

\twocolumngrid

\onecolumngrid

\twocolumngrid
\clearpage
\section{Simulation and Modelling}

\subsection{Cross-Section Estimations}
\label{sec:me_models}

The Snowmass MuIC white paper~\cite{Acosta:2022ejc} considered Higgs production using the \madgraph\ model {\tt sm-no\_b\_mass}, which implements a leading-order (LO) version of the SM Lagrangian and sets the b-quark mass to zero in the parton-level interactions, effectively implementing a "5-flavor scheme." The Higgs Effective Field Theory (HEFT) is also provided in \madgraph, and the \feynrules\ package provides three alternative SM implementations that are stated to be NLO in QCD corrections~\cite{FeynRulesSM}. The latter models come in three forms, denoted {\tt SM\_NLO}, {\tt SM5F\_NLO}, and {\tt SM4F\_NLO} for "\ldots QCD corrections with the top and bottom yukawa [sic] renormalised in the on-shell scheme," "\ldots QCD corrections in the 5 flavour scheme (massless b but non-vanishing bottom yukawa [sic] renormalised in the MSbar)," and "\ldots QCD corrections in the 4 flavour scheme (massive b but bottom yukawa [sic] renormalised in the MSbar)," respectively~\cite{FeynRulesSM}.

For each model, 100,000 events were generated. The PDF used for each was {\tt CT18NNLO}, unless otherwise specified. In all cases, the proton was redefined in \madgraph\ to contain b-quarks as well, e.g. {\tt define p = b b\textasciitilde\ p}. We provide additional details of the simulation in Appendix~\ref{sec:simulation_details}. Results are in Table~\ref{tab:me_calculations}.

Regarding the $tH$ production mechanism it is clear that a \madgraph\ model employing a 5-flavor scheme is essential to more accurately estimating the cross-section in a MuIC. At the level of uncertainty achievable from systematic on the scale (${}^{+0.083\%}_{-1.7\%}$), central scheme (${}^{+0.0\%}_{-1.2\%}$), and PDF (${}^{+19\%}_{-11\%}$) there is no clear difference between {\tt sm-no\_b\_mass} and {\tt SM5F\_NLO-full}. It is apparent that the level of certainty on the b-quark contribution to the proton is a strong influence on the precision with which this production mechanism can be estimated in advance, as PDF uncertainty dominates this calculation.

\onecolumngrid

\begin{table}[htpb]
    \centering
    \begin{tabular}{lccc}
      Model Name   &   \multicolumn{3}{c}{Cross-Section [fb]}\\
                   & $p^+ \mu^- \to H^0 \nu_{\mu} + \textrm{jet}$
                   & $p^+ \mu^- \to H^0 \mu^- + \textrm{jet}$
                   & $p^+ \mu^- \to t H^0 \nu_{\mu}$ \\
      \toprule
       {\tt sm-full} 
            & $65.6$ 
            & $11.5$
            & $0.000117$
            \\
       {\tt sm-no\_b\_mass}  
            & $65.5$
            & $11.7$
            & $0.0240$
            \\
       {\tt heft-full} 
            & $65.5$ 
            & $11.7$
            & N/A
            \\
       {\tt SM\_NLO-full}  
            & $64.0$
            & $11.2$
            & N/A
            \\
       {\tt SM4F\_NLO-full}  
            & $64.0$
            & $11.2$
            & N/A
            \\
       {\tt SM5F\_NLO-full}  
            & $64.0$
            & $11.4$
            & $0.0238$
            \\
       \bottomrule
    \end{tabular}
    \caption{Cross-sections determined using \madgraph\ (\mgamcnlo) and a range of available Standard Model implementations available in the framework or from \feynrules. The PDF set used for all cases is CT18NNLO. In all cases, the pure statistical uncertainly on the estimated cross-section is at the level of $5 \times 10^{-2}~\fb$.}
    \label{tab:me_calculations}
\end{table}
\twocolumngrid

\subsection{Heavy Flavor, PDFs, and Higgs Production}

We investigated the effect on production cross-section resulting from the degree of knowledge of the flavor content of the proton (e.g., at the level of the strange quark and above). Given the role that the bottom quark plays in top-Higgs associated production, the inclusion of heavy flavor in both the underlying model and in the PDF model were of particular interest, should that mechanism be of future interest to a MuIC program.

We compared (Table~\ref{tab:me_calculations_pdfs}) the estimated production cross-sections for the three majors processes --- CC DIS, NC DIS, and top-Higgs associated production --- using a range of PDF sets. We begin by simply comparing the estimates obtained when using the reference PDF set in Ref.~\cite{Acosta:2021qpx} (\texttt{PDF4LHC15\_nlo\_mc\_pdfas}) to our baseline PDF set. We find that the we essentially reproduce the results from that original work. We observe that the top-Higgs associated production is enhanced when utilizing the \texttt{CT18NNLO} PDF set, by about 20\%. We note that, within the estimated theory uncertainties discussed earlier that this difference is not significant.

We also compare the production cross-sections using a foundational PDF fit --- \texttt{CT14}~\cite{Dulat:2015mca} --- that is varied according to a range of assumptions about the fitted number of flavors (\texttt{NF}) and the computation of the PDFs to leading and next-to-leading orders (\texttt{LO} and \texttt{NLO}). We observe no effect on the CC DIS production process. This is compatible with the expectation that this mechanism is dominated by the contributions of valence and the lightest sea quark flavors. This observation is also compatible with the effect on NC DIS Higgs production, which shows negligible changes across the range of assumptions. 

The most interesting pattern emerges from the variation of the number of heavy flavors that can contribute to the proton and the $tH$ mechanism. Since this requires a bottom quark PDF, the variations with \texttt{NF}$<5$ yield no useful predictions (one cannot model a proton containing a bottom quark when there is no PDF to accommodate this hypothesis, and one cannot produce the Higgs this way without the bottom quark). In the case where all six quark flavors are fitted in the PDF, we observe that increasing the order of the computation of the PDF decreases the production cross-section. This observation is statistically valid across the different simulation samples, given the level of uncertainty on the yield of Monte Carlo simulation events. However, in a real data sample this difference would not be observable given the current level of theory uncertainty that contributes to knowledge of this process. That uncertainty would be a significant systematic uncertainty on the interpretation of collider data.

\onecolumngrid

\begin{table}[htpb]
    \begin{center}
    \begin{tabular}{lccc}
      Model Name   &   \multicolumn{3}{c}{Production Process Cross-Section [fb]}\\
                   & CC DIS
                   & NC DIS
                   & $tH$ \\
      \toprule
       {\tt CT18NNLO} 
            & $63.9$ 
            & $11.0$
            & $0.0235$
            \\
       {\tt PDF4LHC15\_nlo\_mc\_pdfas}  
            & $63.1$
            & $10.9$
            & $0.0196$
            \\
       {\tt CT14lo}  
            & $63.6$
            & $11.2$
            & $0.0276$
            \\
       {\tt CT14lo\_NF3}  
            & $63.8$
            & $11.0$
            & N/A
            \\
       {\tt CT14NLO\_NF3} 
            & $63.7$ 
            & $10.8$
            & N/A
            \\
       {\tt  CT14lo\_NF4}  
            & $64.0$
            & $11.1$
            & N/A
            \\
       {\tt CT14NLO\_NF4}  
            & $63.9$
            & $11.2$
            & N/A
            \\
       {\tt CT14lo\_NF6}  
            & $63.6$
            & $11.2$
            & $0.0273$
            \\
       {\tt CT14NLO\_NF6}  
            & $63.6$
            & $11.4$
            & $0.0190$
            \\
       \bottomrule
    \end{tabular}
    \end{center}
    \caption{Cross-sections determined using \mgamcnlo, the Standard Model implementation \texttt{SM5F\_Hcc\_NLO} available in the framework, and a range of PDF sets at different orders and with different quark flavor assumptions. In all cases, the pure statistical uncertainly on the estimated cross-section is at the level of $2 \times 10^{-5}~\fb$.}
    \label{tab:me_calculations_pdfs}
\end{table}

\twocolumngrid

\subsection{Background Processes}
\label{sec:backgrounds}

It is expected that there will be at least a few key background processes in the search for $\Hcc$:

\begin{enumerate}
    \item Inclusive dijet production: while heavy flavor will be part of this, given charm and bottom jet mis-identification rates it is expected that the much larger production cross-section for inclusive dijets at a MuIC will lead to contamination of the final sample of \Hcc\ candidates, even if light jet suppression is very good for such algorithms. This should be reducible by increased attention to light-jet rejection performance in any jet identification algorithm.
    
    \item Charm dijet production: heavy flavor dijet production will, of course, directly contaminate the \Hcc\ sample. This should be reducible by using the kinematics of Higgs decay to discriminate from this population of events, though of course this population will contain real charm jets.
    
    \item \Zcc\ events: production of the Z boson followed by decay to charm is expected to be a main, semi-irreducible background. It's reducible in the sense that the Z dijet mass will be different from the Higgs dijet mass, but given dijet mass resolution at the level of tens of \gev\ the key to separating these classes of events will depend on detector resolution and jet calibration. A \mgamcnlo\ simulation of this process yields an expected inclusive $Z$ production cross-section of $0.331\pb$ ($3.0$) for CC DIS (NC DIS).
\end{enumerate}

It is possible to simulate aspects of the above backgrounds in \madgraph\ and \pythia. Feynman diagrams for the processes utilized in \mgamcnlo\ are shown in Appendices~\ref{sec:ncdis_dicharm_mg5} and \ref{sec:ccdis_dicharm_mg5}. However, it is challenging to simulate all of these background fully inclusively, as most of the events produced by these underlying mechanisms (especially the non-resonant dijet backgrounds) tend to result in events with extremely soft jets or other factors that make them trivially reducible (e.g. the detector will not see many or most of the products of the collision).

\onecolumngrid

\begin{figure}[tbp]
    \centering
    \includegraphics[width=0.45\linewidth]{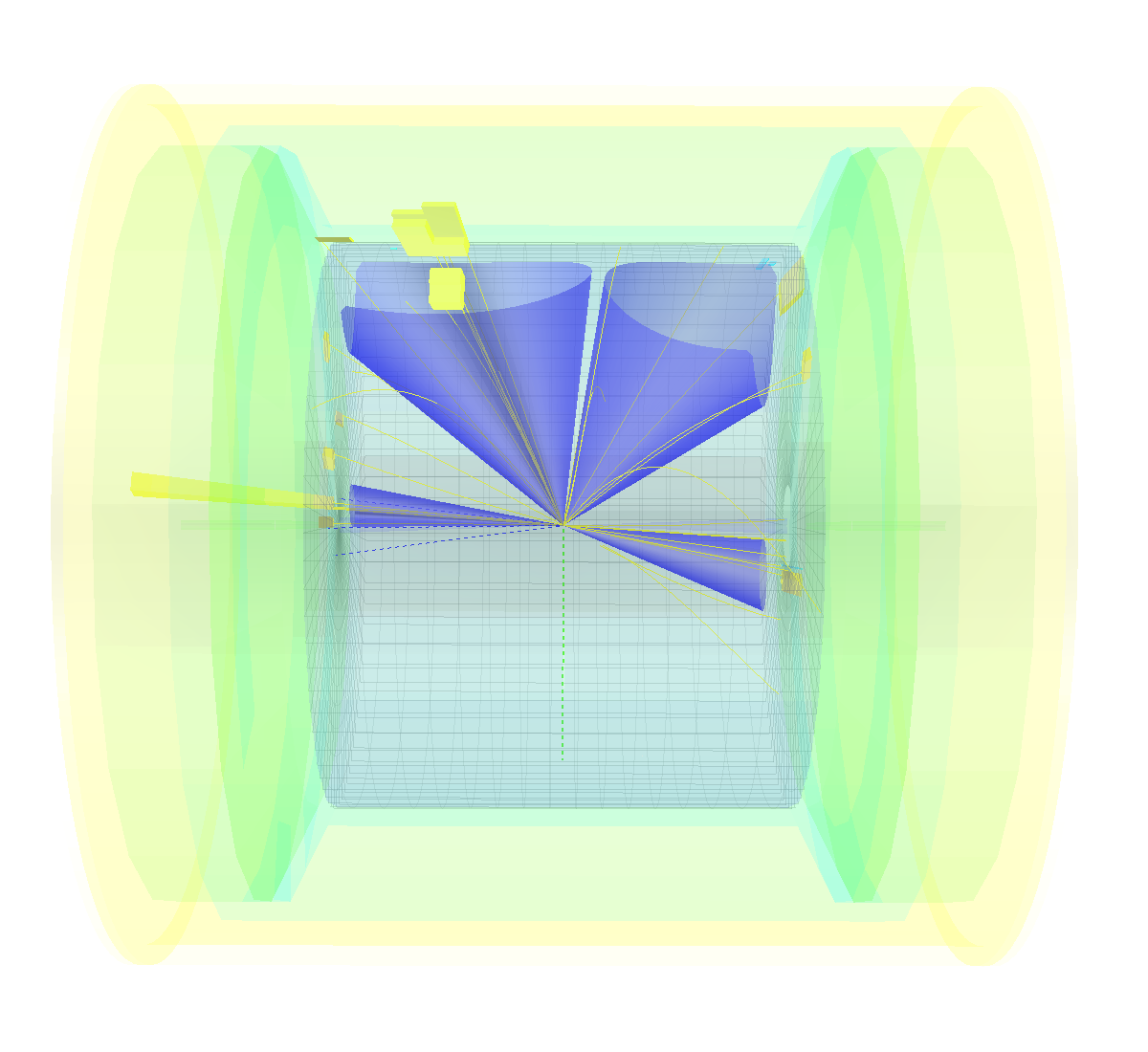}
    \includegraphics[width=0.45\linewidth]{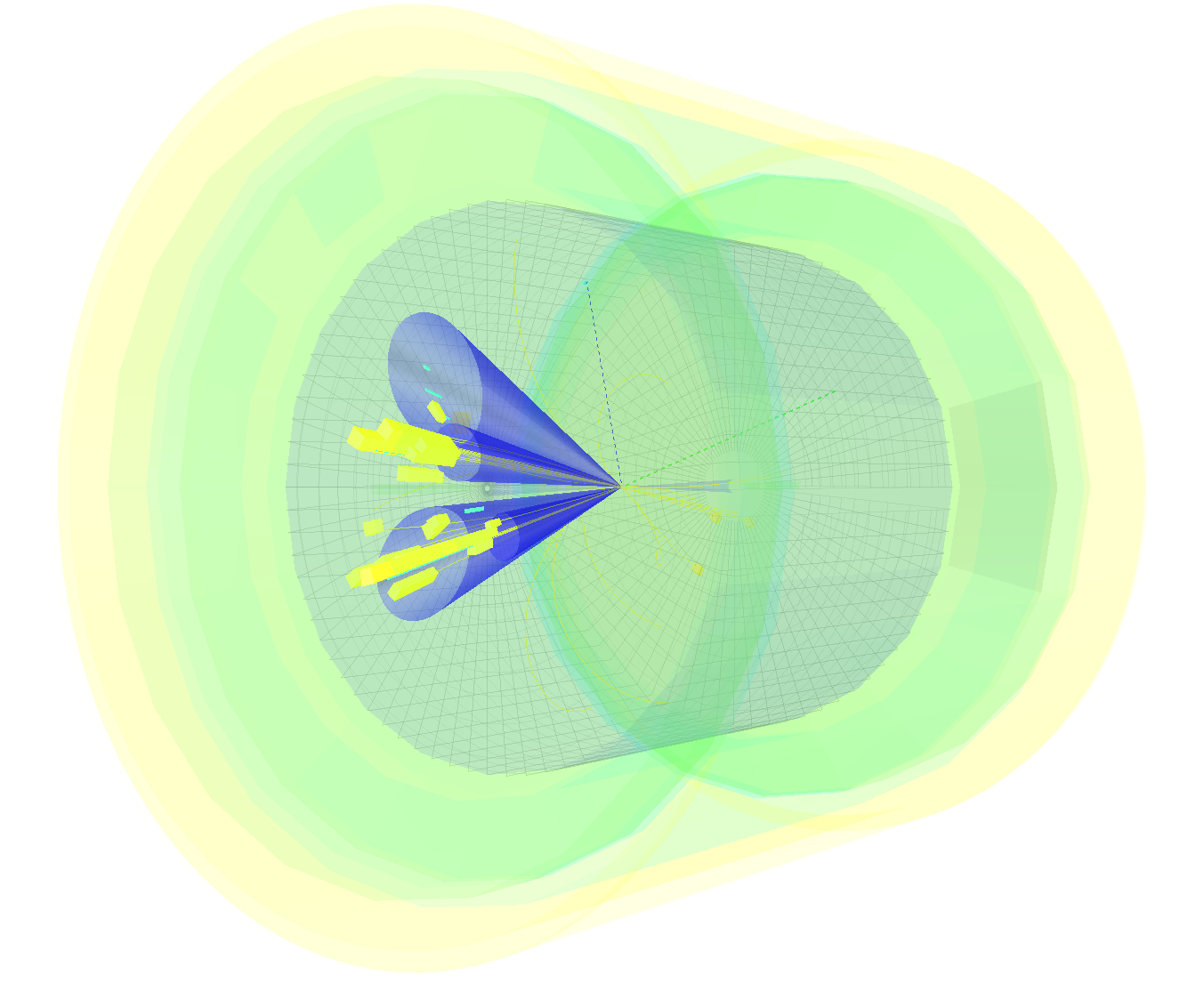}
    \caption{Event displays of a CC DIS \Hcc\ event (left) and an NC DIS di-charm event (right), both of which pass all prototype selection criteria described in Section~\ref{sec:analysis}. Although the perspectives are slightly different, the muon-beam (backward) direction is on the left-most side of each image. In the signal event, the backward two jets (blue cones) are associated with the Higgs decay. In the background event, the two most backward jets were associated with the candidate Higgs when none is present. No reconstructed muons are present in either event.}
    \label{fig:event_displays}
\end{figure}

\twocolumngrid

We adopt two approaches to simulating the di-charm background. The first is almost fully inclusive, with only a generator-level restriction that a final state muon satisfy $|\eta| < 6$ (only in NC DIS) and that there be at least one final-state quark where $p_q>5\gev$ (all production mechanisms). We refer to this as "inclusive di-charm." A second sample is generated also requiring that $m_{cc} > 50~\gev$, similar to Ref.~\cite{Acosta:2022ejc}. We refer to this as "high-mass di-charm." Inclusive di-charm events are essential in revealing the characteristics in missing energy, muon, and jet kinematics that can be used as a first approach to rejecting di-charm events. The high-mass di-charm simulation is essential for a higher statistics sample of this background more likely to pass final selection criteria. Of course, as we restrict the phase space at production we incur the cost of larger theoretical uncertainties. We will return to the challenges of the di-charm background at the conclusion of this note. We estimate the inclusive (high-mass) di-charm production cross-section to be $509,000~\fb$ ($97,700~\fb$). We simulate 200,000 collisions for each of these samples, a yield generally at or far below what would be expected for our target integrated luminosities (e.g., $1\invfb$ or $100\invfb$).

The systematic uncertainties due to scale, $\alpha_s$, and PDFs  are generally larger than these statistical effects (which can be reduced merely by simulating more events of this type). For example, the scale uncertainty is about 20\% while the central scheme uncertainty is over 100\% in both cases described above and worsens as one further restricts the phase space of the process. As expected, the more exclusive the process the larger are some associated theory uncertainties. Restricting the simulation from the fully inclusive approach (e.g. no limitations on jet \pt\ or $\eta$) at this level nonetheless retains a sample of events that could be reconstructed by a detector, albeit in a phase space that is only part of the total inclusive rate. 

The graphs in this note are generally normalized in one of two ways. When "density" appears on the y-axis, the contents of the figure have been normalized using the area of distribution and the width of the histogram bins. Otherwise, we normalize the samples to their production cross-sections (and branching fractions, where appropriate) and a target luminosity of 100~\invfb. The density normalization allows for a comparison of shape differences in samples and populations; the luminosity normalization allows for a sense of how these populations will look in a future data sample.

We note that, unlike in the case of $\Hbb$ explicitly studied in Ref.~\cite{Acosta:2021qpx}, $\Hcc$ appears to suffer from at least two major disadvantages at the level of a generator- or pre-selection of events:
\begin{enumerate}
    \item The branching fraction for $\Hcc$ is significantly lower than for $\Hbb$ owing to the much smaller Yukawa coupling to charm quarks (a factor of about 20);
    \item The probability of non-resonantly pair producing charm quarks in a high-energy collider is much greater than that of producing a pair of bottom quarks, due to the much smaller mass of the charm quark.
\end{enumerate}
These two factors could make the study of $\Hcc$ particularly challenging, even in this kind of collider environment. We explore the reconstruction and selection of these events to identify the basic challenges of this process after a modest, baseline selection.

\subsubsection{Visualization of Events}

We show \delphes\ event displays of the CC DIS \Hcc\ process and for the NC DIS di-charm process in Fig.~\ref{fig:event_displays}. The events shown here pass all selection criteria outlined in Section~\ref{sec:analysis} and are considered extremely "signal-like." The \Hcc\ event has two backward jets that are identified as charm-induced and form a di-jet invariant mass consistent with the Higgs particle. Other jets are present as a result of the proton beam remnant, radiation, etc. The background event is induced by a NC DIS process but no reconstructed muon is present. The two most backward jets are the ones associated with the Higgs candidate. Both events have significant missing transverse momentum. The analysis we outline later is very basic, but these event displays may hint at topological differences in jets that may be exploited by more advanced methods.

\subsection{Jet Reconstruction in Detector Model}
\label{sec:jet_reco}

The detector model employed in this study is based on the "EIC Generic 3T" model provided in the {\tt delphes\_eic} package~\cite{Arratia:2021uqr}. The model is modified, however, in a few key ways:

\begin{itemize}
    \item The calorimeter design can be changed to try alternatives, including the original EIC Generic calorimeter model~\cite{AbdulKhalek:2021gbh}, a model similar to that proposed for the ATHENA detector concept~\cite{ATHENA:2022hxb}, and a model similar to the CMS calorimeter adopted in Ref.~\cite{Acosta:2022ejc}. These are referred to as the "EIC Generic," "ATHENA," and "Snowmass" concepts, respectively. We generally adopt the "Snowmass" configuration in our studies. While it does not appear to provide the best resolution, it provides a somewhat more conservative baseline. All calorimeter models include an electromagnetic and hadronic calorimeter component with identical coverage in pseudorapidity, though not identical cell granularity.
    \item Jets can be reconstructed using only calorimeter towers ("Calorimeter-only", or "CaloJets") or using particle flow. We allow for both kinds of jet and missing energy reconstruction to facilitate studies of the effect of particle flow algorithms on a future Higgs candidate reconstruction (simply referred to henceforth as "Jets"). Our baseline, unless otherwise stated, uses particle flow (c.f. Ref.~\cite{CMS:2009nxa}). All jet reconstruction approaches use the anti-$k_T$~\cite{Cacciari:2008gp} algorithm.
    \item A muon efficiency model is added to represent the performance of a future, low-angle muon spectrometer system (intended primarily for beam muon reconstruction in, for example, neutral-current DIS).
\end{itemize}

\subsubsection{Jet Parameter Optimization}
\label{sec:jet_parameter_opt}

LHC experiments employ a range of $R$-parameter choices for anti-$k_T$ jets. For example, standard jets in CMS or ATLAS are typically defined with $R$ parameters of 0.5 or 0.4, respectively~\cite{CMS:2016lmd,ATLAS:2020cli}. Highly Lorentz-boosted heavy objects decaying to jets will result in larger radius jet clusters, and the typical $R$ parameter for this situation is 1.0~\cite{ATLAS:2020gwe}. Substructure within a large-radius jet can be determined by re-running jet clustering inside the object using charged particle tracks, and typically this is done using jets with a narrower $R$ parameter (e.g. 0.2) or a sliding scale that adjusts the $R$ parameter depending on the kinematics of the substructure jet relative to the parent jet.  It is not \textit{a priori} obvious what is the right or best choice for jet reconstruction of $\Hcc$ at a future MuIC.

\begin{figure}[htbp]
    \centering
    \includegraphics[width=0.9\linewidth]{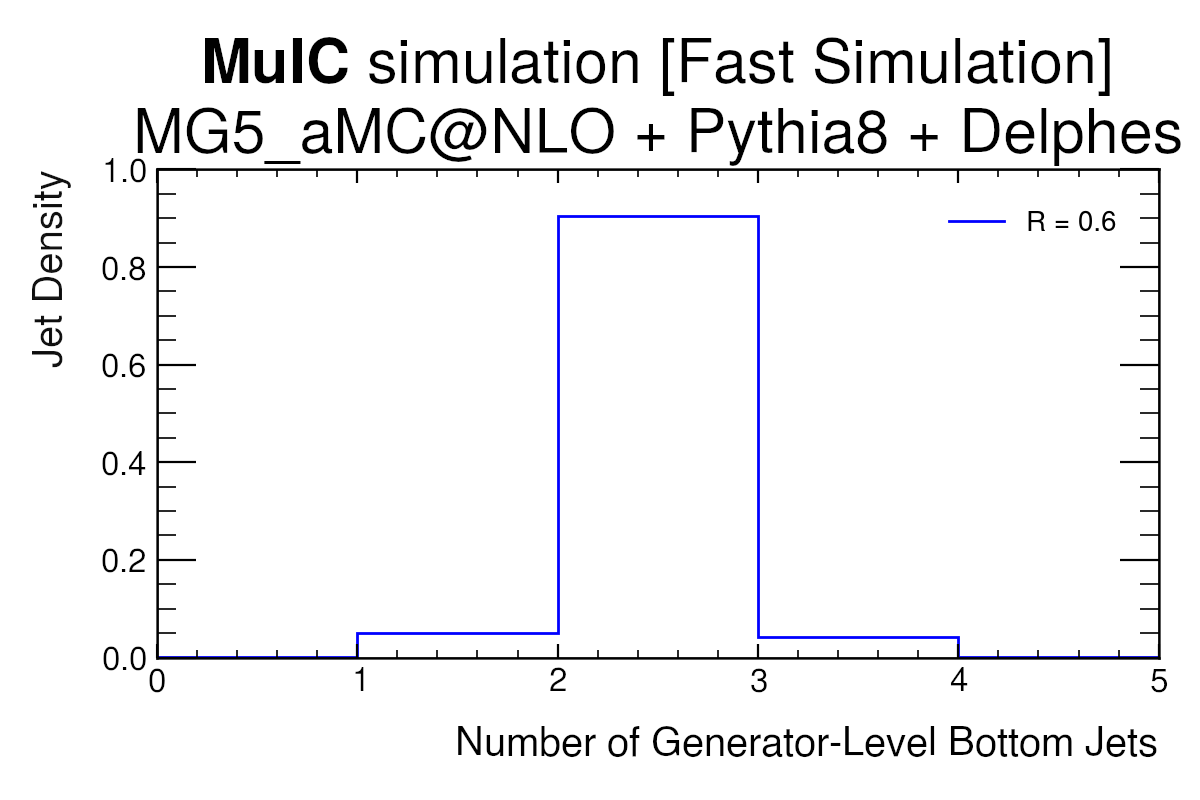}
    
    \includegraphics[width=0.9\linewidth]{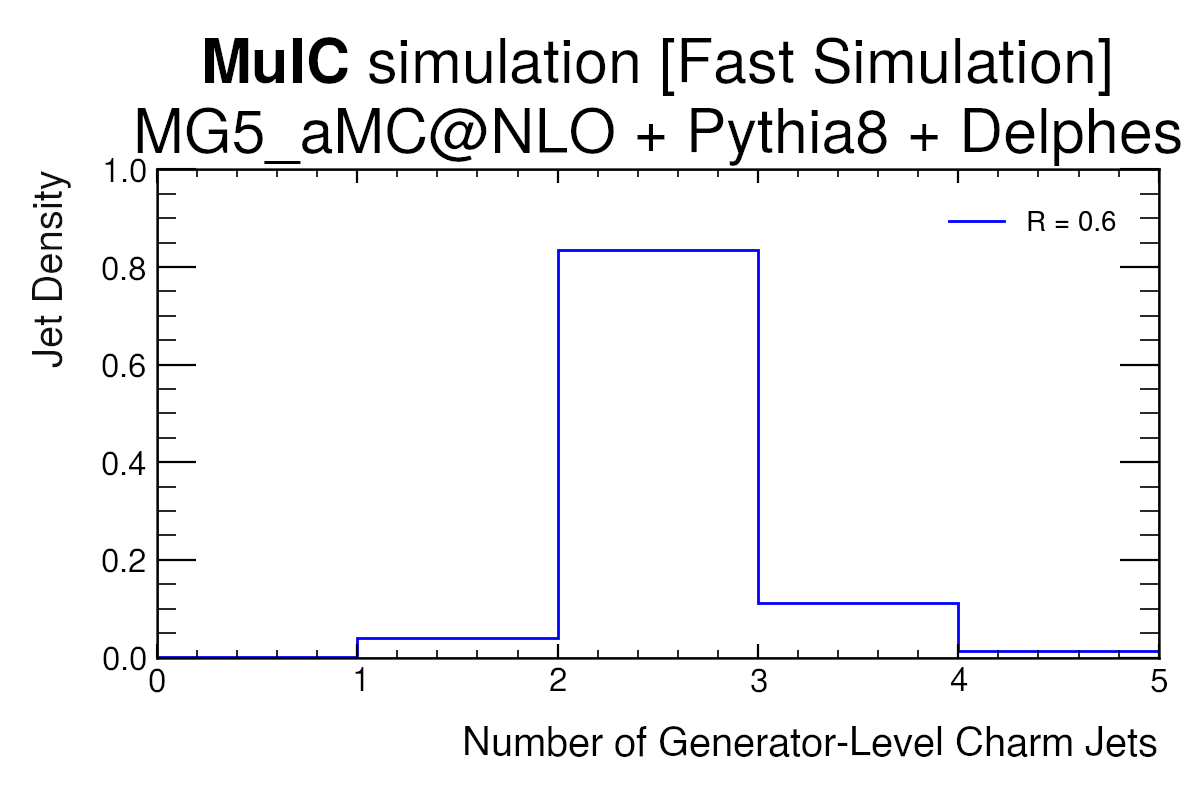}
    \caption{Number of bottom (upper) and charm (lower) quarks produced from a single collision event. All events are derived from either \Hbb\ and \Hcc\ simulation, respectively.}
    \label{fig:r6_events}
\end{figure}

We reconstructed jets using a few different approaches and selected an $R$ parameter that balanced Higgs mass accuracy and resolution. We scanned across the possibilities $R = 0.4-1.0$ in steps of $0.2$. To avoid over-constraining the choice to just one possible heavy flavor, we looked at the quality of both di-charm and di-bottom reconstruction. Results are shown in Fig. ~\ref{fig:r6_events}.

We hypothesized that a baseline means of selecting the best $R$ parameter was to simply observe the rate at which $\Hbb$ and $\Hcc$ simulated events were reconstructed as having two heavy flavor jets. We recognize that the production of charm quarks in the proton remnant affects this hypothesis, as a fraction (at the level of a few percent) of $\Hcc$ events will contain at least 3 charm jets. Nevertheless, most events would be expected to contain only two such heavy flavor jets.

\begin{figure}[htbp]
    \centering
    \includegraphics[width=0.9\linewidth]{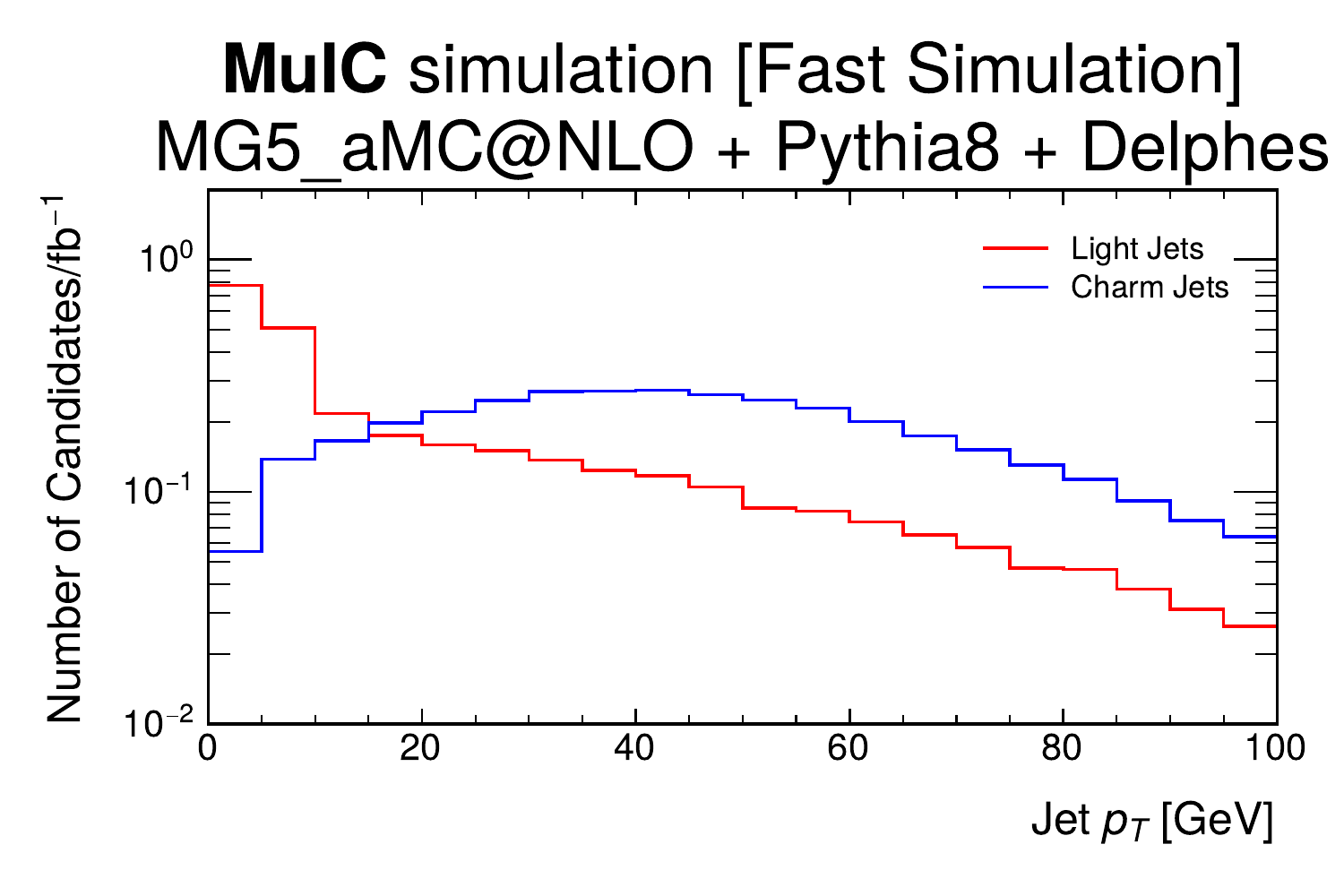}
    \caption{Jet \pt\ spectrum is shown from CC DIS \Hcc\ simulation, distinguishing true charm-initiated jets and other kinds of jets ("light jets").}
    \label{fig:jet_spectrum}
\end{figure}

For iterations $R = 0.4$, $R = 0.6$, $R = 0.8$, and $R = 1.0$, events with two b-jets occurred with a frequency of $87\%$, $90\%$, $91\%$, and $91\%$, respectively. The statistical uncertainty on these rates is smaller than the reported precision. Di-charm jet production occurred under the same parameters with frequencies of $82\%$, $84\%$, $84\%$, and $85\%$. We observe that for  both heavy flavor possibilities there is a clear few percent-level rise in the rate of di-jet reconstruction as we move from $R=0.4$ to $0.6$. However, after that, the rise is less dramatic and we generally observe that the reconstruction rate of heavy flavor di-jets is flatter for larger jet $R$ parameters. We select a point on the trend that is past the early rise and early in the plateau. Thus,  $R = 0.6$ was chosen as the most optimal $R$ value for all future work in this note. A comparison of light and charm jets in CC DIS \Hcc\ simulation is shown in Fig.~\ref{fig:jet_spectrum}.

\subsubsection{Jet Resolution}
\label{sec:jet_resolution}

In the context of our selected jet reconstruction, we can explore the fast-simulation-level expected quality of jets used in the reconstruction of Higgs particles. In particular, the resolution of a jet variable $X$ is defined as $(X_{reco} - X_{true})/X_{true}$, where "reco" labels the variable determined purely from a reconstruction-based approach and "true" labels the variable as determined from a Monte Carlo generator-level set of information. Generator-level jets are defined as jets reconstructed just as they would be from calorimeter information, but instead employing all long-lived stable particles from the simulation before detector-level effects are simulated.

\begin{figure}[htbp]
    \centering
    \includegraphics[width=0.9\linewidth]{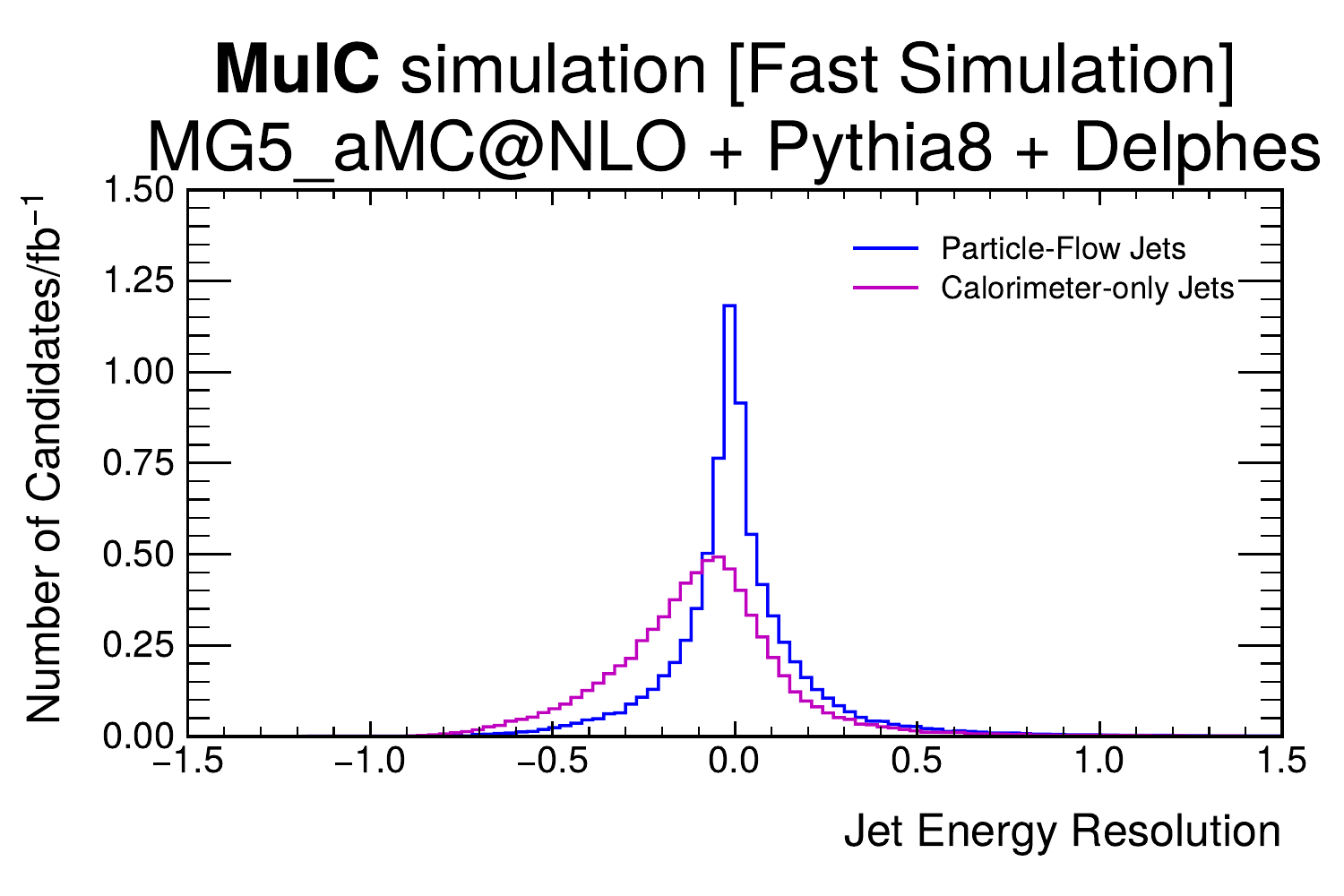}
    \caption{Jet energy resolution is shown for calorimeter-only jets and particle flow-based jets.}
    \label{fig:jet_ereso}
\end{figure}

The typical jet energy resolution at a modern experiment, such at ATLAS, varies with the kinematics (e.g. $\pt$) of the jet object. ATLAS obtains a jet energy resolution of about 24\% at $\pt=20~\gev$ that improves to about 6\% at $300~\gev$. The CC DIS production of \Hcc\ in our simulation results in charm jets with a typical \pt in the range of $30-40~\gev$. The energy resolution obtained in our fast simulation is shown in Fig.~\ref{fig:jet_ereso}. We estimate it for both CaloJets and Jets. We can clearly see the positive effect of employing particle flow, as this improves significantly the resolution of the energy and its accuracy to the true value from the generator-level jet. The root mean-square (RMS) of the particle flow jet energy resolution is about 21\%, consistent with typical LHC jet energy resolution measurements for a similar \pt\ spectrum.

\subsection{Higgs Jets and Kinematics}
\label{sec:higgs_jets}

We compare the \pt, azimuthal angle ($\phi$), and pseudorapidity ($\eta$) between generator-level jets and reconstructed jets. As a basic check, we observed that the distribution of $\phi$ was  consistently isotropic, as expected given both a symmetric detector model and a zero crossing angle for the muon and proton beams.

\begin{figure}[htbp]
    \centering
    \includegraphics[width=0.9\linewidth]{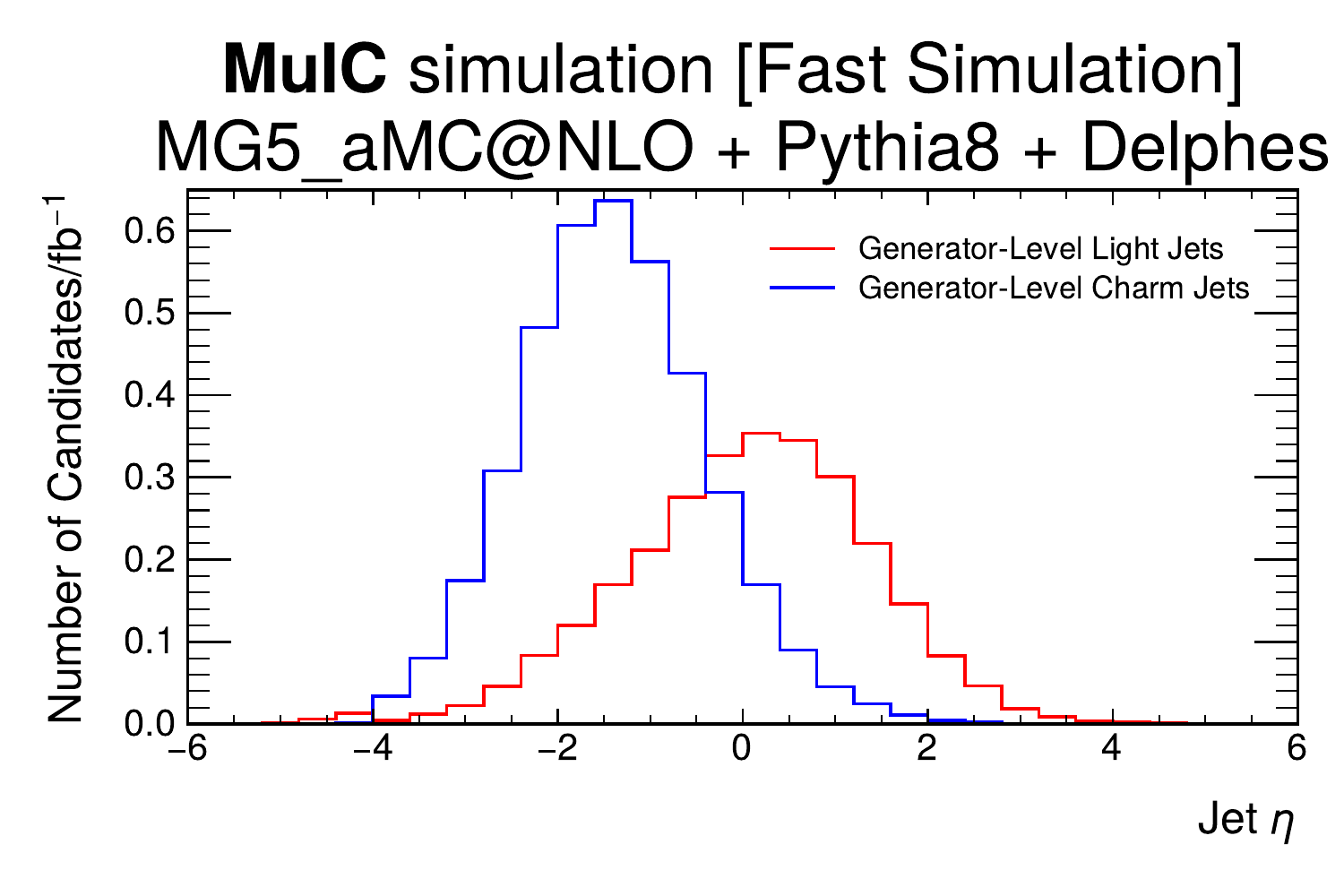}
    \caption{$\eta$ distribution is shown for generator-level jets from \Hcc\ simulation subdivided into charm and light-jet populations.}
    \label{fig:eta_plots}
\end{figure}

We observe that the jets in \Hcc\ CC DIS events are typically produced in the backward (muon beam) direction, which would be expected from the asymmetric beam kinematics (Fig.~\ref{fig:eta_plots}). The proton remnant would be expected to travel in the forward direction; indeed, separating charm jets from light jets, we observe that the light-jet component of the $\eta$ distribution more often contains a population of jets in the forward direction. Charm jets, primarily resulting from Higgs decay, tend to travel forward. The jet density in $\eta$ peaks around $\eta = -1.5$. 

\begin{figure}[htbp]
    \centering
    \includegraphics[width=0.45\linewidth]{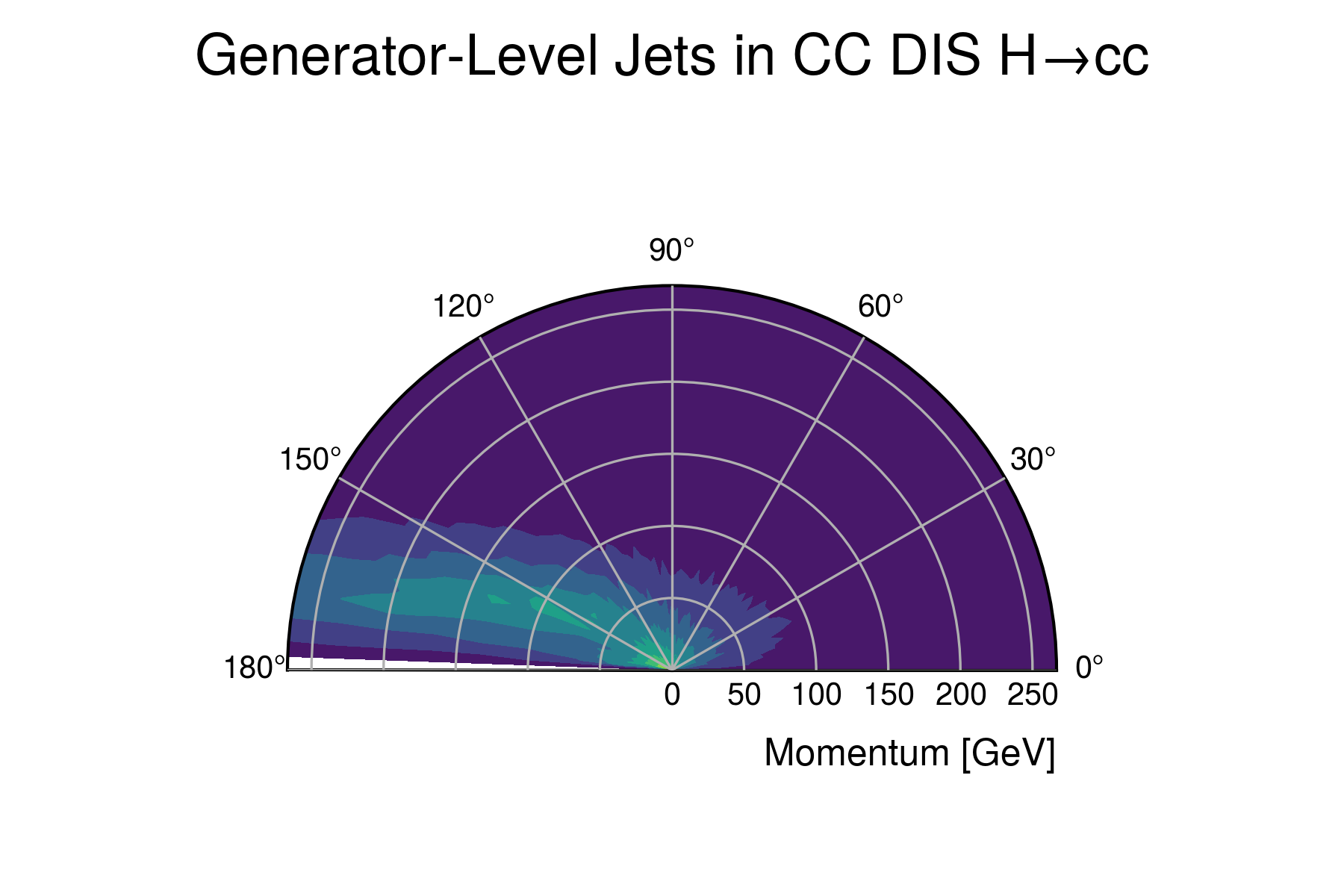}
    \includegraphics[width=0.45\linewidth]{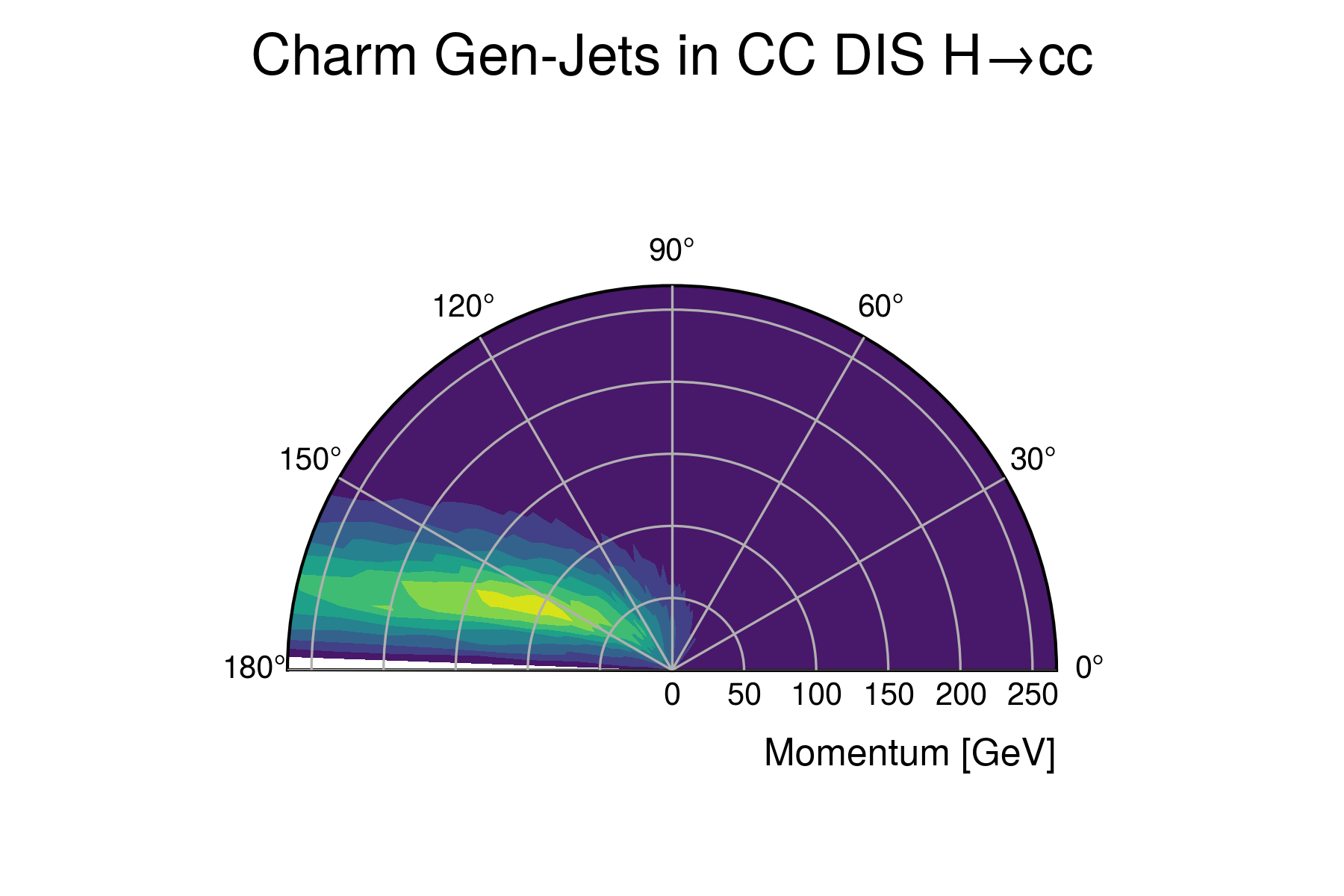}
    
    \caption{Coverage plot for all generator-level jets (top) and for generator-level charm jets (bottom) illustrating the relationship between the jet momentum $p$ as a function of polar angle $\theta$.}
    \label{fig:charm_coverage_plot}
\end{figure}

Tracking and calorimetry in a future MuIC detector will need to provide sufficient coverage and pixellation in the backward direction. We show momentum and polar angle coverage plots in Fig.~\ref{fig:charm_coverage_plot}, emphasizing the typical region populated by generator-level jets in a collider of this configuration (275x960~$\gev^2$  protons on muons). The majority of decay products fall at $\eta = [-2.5, 0]$. Taking this into account with the specified jet parameter $R = 0.6$, a MuIC collider detector will require calorimeter and tracking coverage out to at least $|\eta| = 4.0$, with emphasis for jet reconstruction in the direction of the muon beam.

\onecolumngrid

\twocolumngrid

\begin{figure}[htbp]
    \centering
    \includegraphics[width=0.9\linewidth]{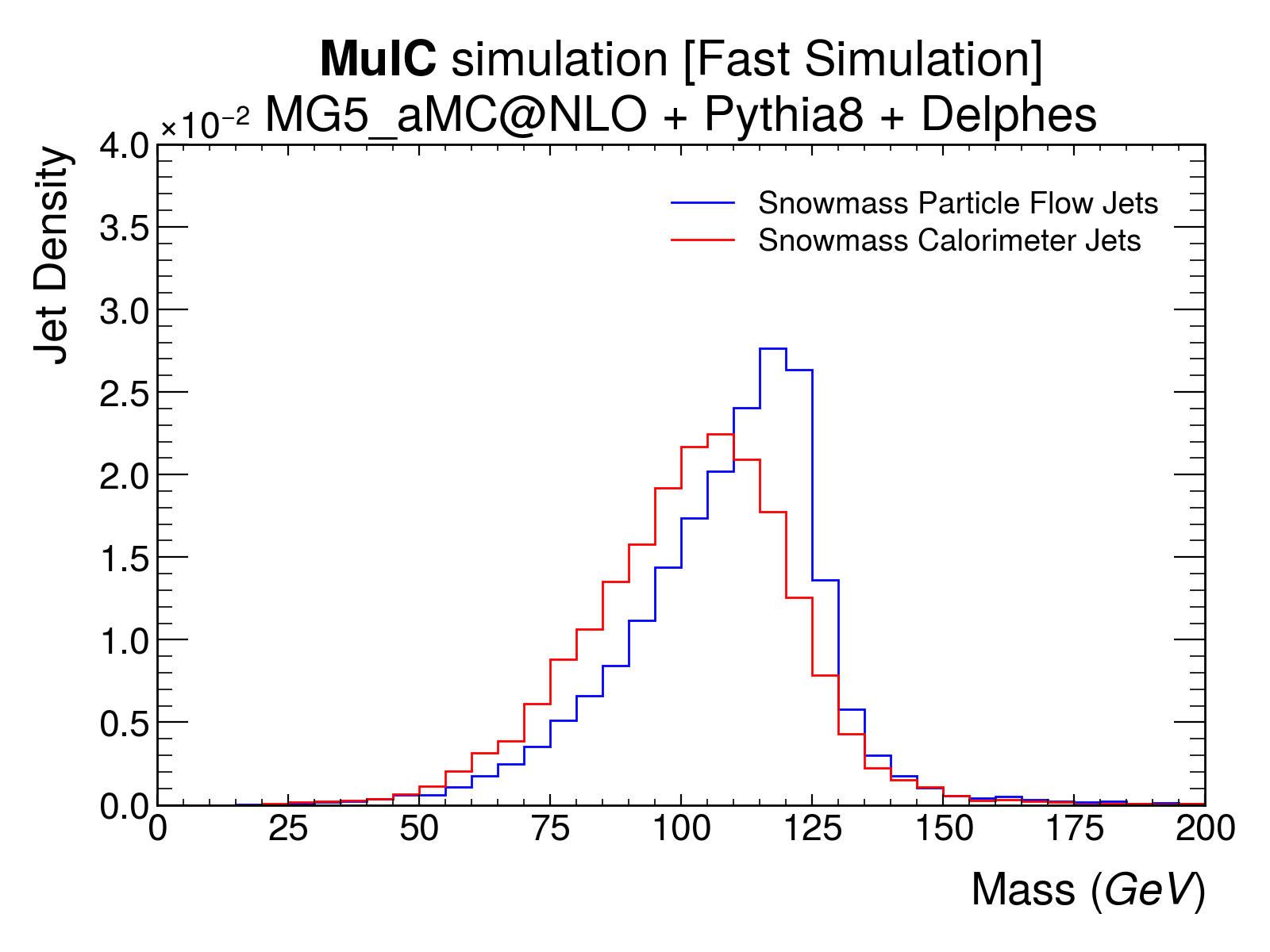}
    \caption{Comparison of Particle Flow and Calorimeter Jets for $H \to c\bar{c}$ Higgs reconstruction.}
    \label{fig:pflow_calo_comparison}
\end{figure}

We have observed that the jet energy resolution for charm jets is consistent, in this fast simulation, with typical resolutions at existing experiments. We investigate also the mass reconstruction of the dijet system. This is expected to be a key discriminant in any Higgs reconstruction. When comparing jets and CaloJets (Fig.~\ref{fig:pflow_calo_comparison}) it is clear that particle flow provides a more accurate and precise mass reconstruction. 

Our baseline configuration results in a median (peak) dijet mass of $111~\gev$ ($117~\gev$ for truth-matched charm jets in the CC DIS \Hcc\ simulation. The standard deviation of the mass peak is about $21~\gev$. We would expect to thus be able to separate the \Hcc\ mass peak from a $Z \to \ccbar$ background mass peak at the level of about $1~\sigma$; however, this is without employing standard methods for improving the calibration of jets, such as jet-energy scale and muon-in-jet corrections (which compensate for jet mis-reconstruction and energy loss due to semi-leptonic heavy quark decays). These results are consistent with those obtained using $H \to b\bar{b}$ decay in Ref.~\cite{Acosta:2022ejc}.

\begin{figure}[htbp]
    \centering
    \includegraphics[width=0.9\linewidth]{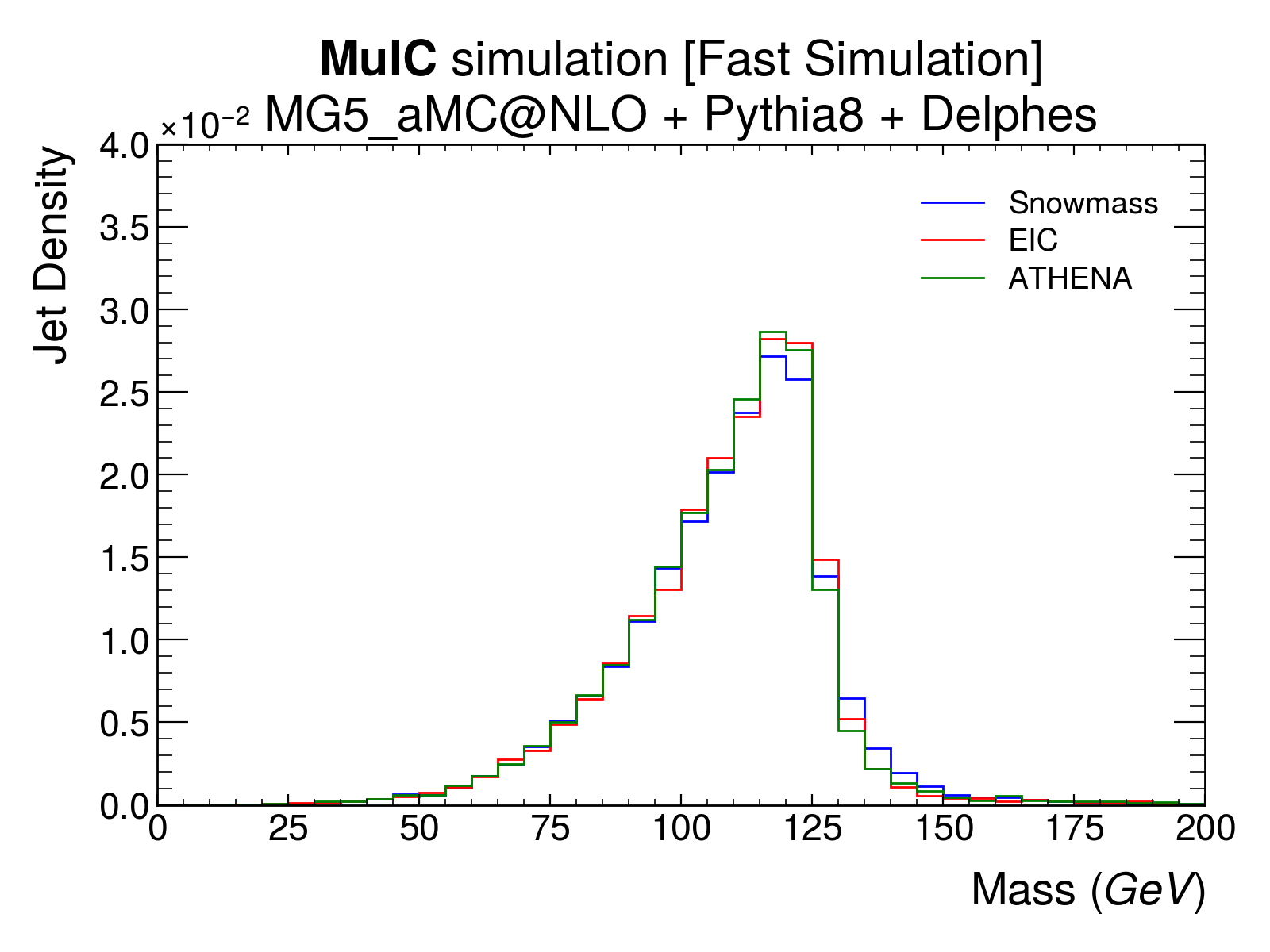}
    \caption{Dijet invariant mass distribution for CCDIS $H \to c\bar{c}$ compared between alternative calorimeter models.}
    \label{fig:h_to_cc_mass}
\end{figure}

We also compare the mass reconstruction using the different calorimeter models described earlier. We see no particular advantage in this particular quantity in choosing one calorimeter model over the others (Fig.~\ref{fig:pflow_calo_comparison}). The importance of calorimeter granularity and coverage will be more apparant in the study of missing momentum.

We adopt a charm-tagging performance for jets similar to that in Ref.~\cite{ATLAS:2022ers}, with 27\%, 8\%, and 1.6\% c-jet, b-jet, and light-jet efficiency, respectively. This is taken as independent of the jet momentum, though in reality we recognize that there would be angular and momentum effects that would need to be folded into the development and performance assessment of any charm-tagging algorithm.

\subsection{Muon Identification}
\label{sec:muon_id}

The production of di-charm final states, from either resonant or non-resonant mechanisms, will occur more often in NC DIS than CC DIS. However, Higgs production is dominated by the CC DIS mechanism. Rejecting the significant production of charm through NC DIS will require attention to muon and missing momentum reconstruction.

\begin{figure}[htbp]
    \centering
    \includegraphics[width=\linewidth]{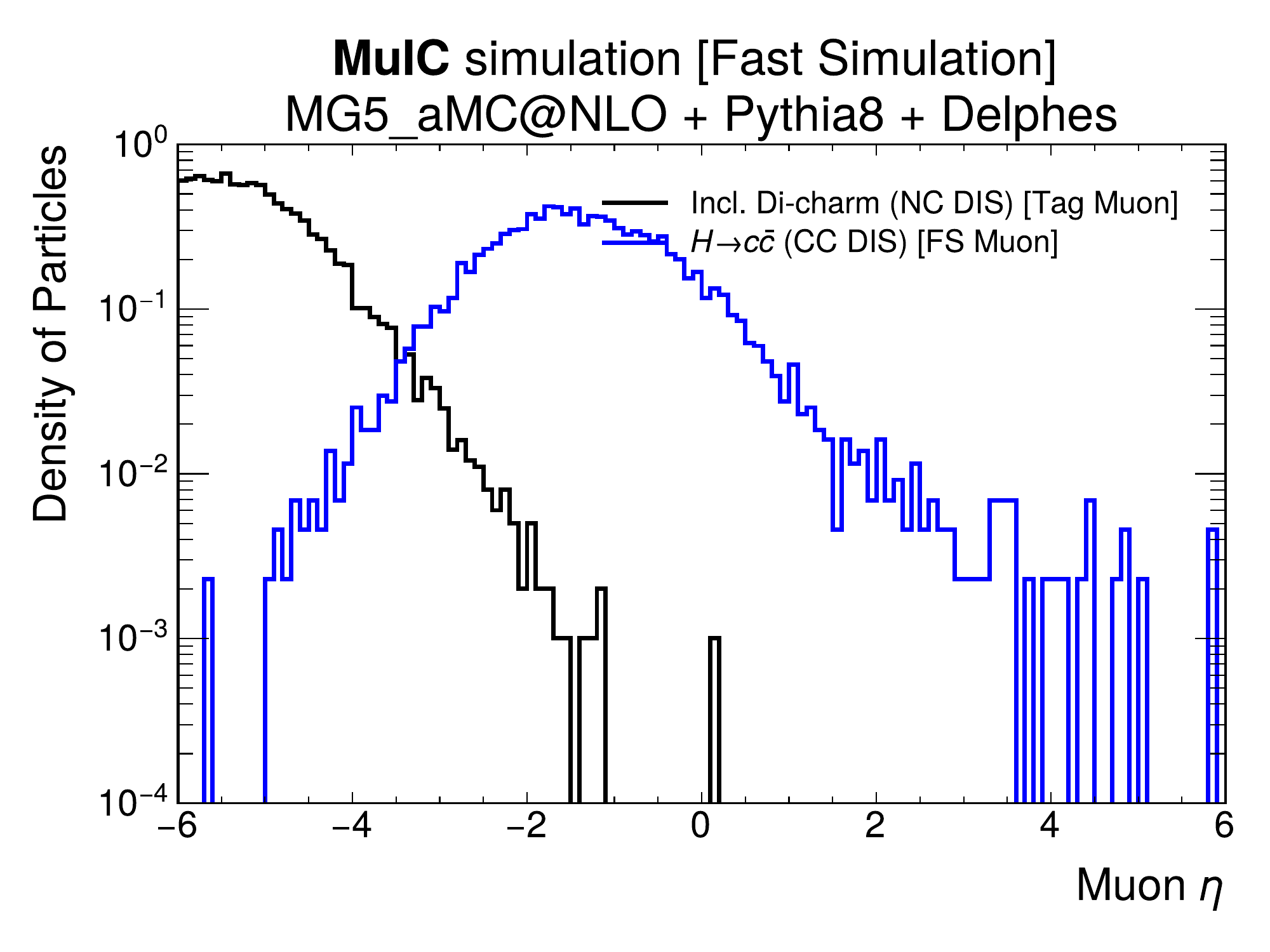}\\
    \includegraphics[width=\linewidth]{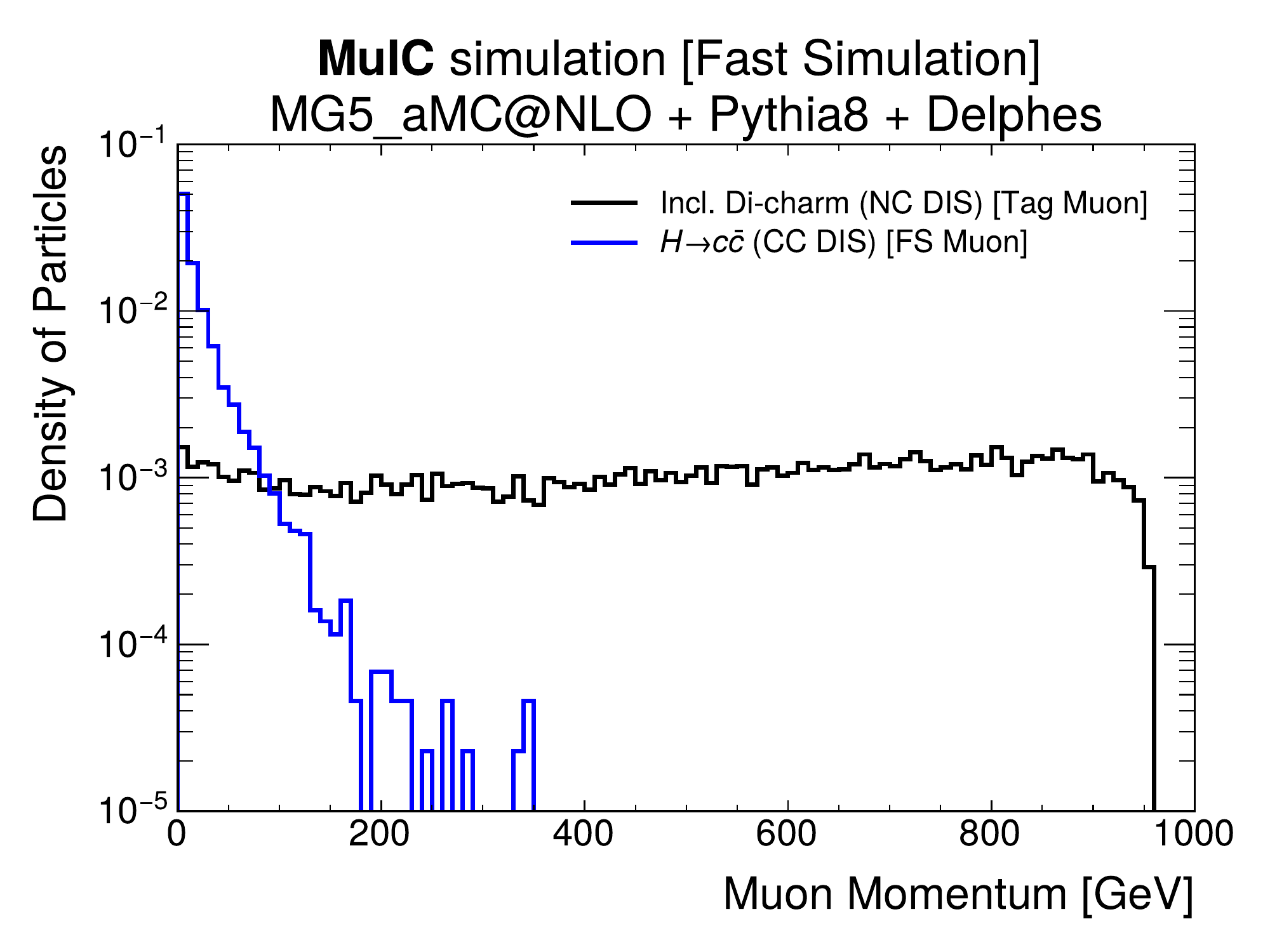}
    \caption{Generator-level characteristics of muons from the beam-beam interaction (tagging) or from final-state decays of heavy particles (FS).}
    \label{fig:muon_tag}
\end{figure}

A backward muon with high energy will be a potential hallmark of the NC DIS process. At the generator level, muon momentum and pseudorapidity both show the characteristics expected of these "tagging muons" --- those scattered from the original beam-beam interaction (Fig.~\ref{fig:muon_tag}). Tagging muons from NC DIS interactions are characteristically in the direction of the muon beam ($\eta < -3$) and very high in momentum, typically with more than 10\% of the lepton beam's original energy. Final-state muons, resulting from the decay of heavy particles, have a much softer momentum spectrum and tend to be much more central in the detector (following more the angular distribution of jets).

Identifying these muons will require the deployment of a far-backward muon system that can detect high-momentum minimum ionizing particles all the way back to $\eta \approx 6$ or lower. LHCb is an excellent example of an experiment with a low-angle muon spectrometer that could serve as a model for a future MuIC detector subsystem with this capability~\cite{Alves:2012ey,Anderlini:2020ucv}. We assume a conservative LHCb-like  performance in the region $\eta=[-5.0,4.0]$~\cite{LHCbPerformance}. This results in 97\% (3\%) muon (pion) efficiency throughout the region of acceptance for muons.

\subsection{Missing Momentum and Energy in CC DIS}
\label{sec:met}

\begin{figure}[htbp]

    \includegraphics[width=0.9\linewidth]{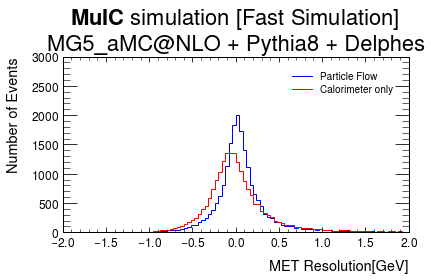}
    \caption{MET Resolution for calorimeter only and particle flow. Events are from CC DIS signal simulation. }
    \label{fig:met_resolution}
\end{figure}

Missing momentum and energy are expected to be key discriminators in the study of $\Hcc$ decay, especially in distinguishing CC and NC DIS production. Missing energy reconstruction in \delphes\ is performed using the energy flow objects (tracks and clusters). Missing transverse four-momentum is computed simply using
\begin{equation}
    \vec{p}^{miss}_T = - \sum_i \vec{p}_i 
\end{equation}
where $\vec{p}_i$ is the momentum of an energy flow object. An example of the baseline resolution ($(p^{miss,reco}_T - p^{miss,true}_T)/p^{miss,true}_T$ is shown in Fig.~\ref{fig:met_resolution}, comparing \met\ calculated using raw calorimeter objects and using particle flow objects. As expected, particle flow-based reconstruction provides a superior resolution and fidelity to the true missing transverse momentum.

\begin{figure}[htbp]
    \centering
    \includegraphics[width=0.9\linewidth]{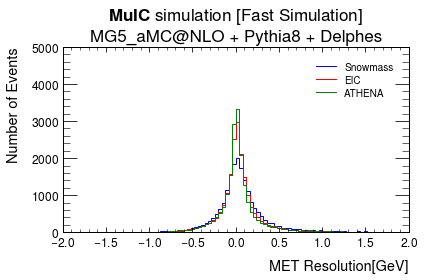}
    \caption{Comparison of the MET Resolution for the different calorimeter models in $\Hcc$ deay.}
    \label{fig:met_resolution_calomodels}
\end{figure}

As an assessment for future consideration of different calorimeter granularity designs, we also calculate the \met\ for the alternative calorimeter models. A comparison in CC DIS signal simulation is shown in Fig.~\ref{fig:met_resolution_calomodels}. These include the ATHENA-like model, a baseline EIC detector model, and what we refer to as the Snowmass calorimeter (our baseline for this note). Only particle flow is utilized. We observe that the Snowmass granularity and coverage offers the overall largest resolution of the three designs. For example, in this fast simulation the Snowmass particle-flow-based \met\ resolution, determined by the standard deviation of the distribution, is about 33\%. In comparison, the resolution in the EIC and ATHENA models (which are similar in granularity) is about 29\%. This suggests both that our results from this note will be conservative and that designs with more granularity would benefit future experiments that need stronger \met\ resolution.

\subsection{Missing Momentum and Energy in NC DIS}
\label{sec:met_ncdis}   

\begin{figure}[htbp]
    \centering
    \includegraphics[width=0.9\linewidth]{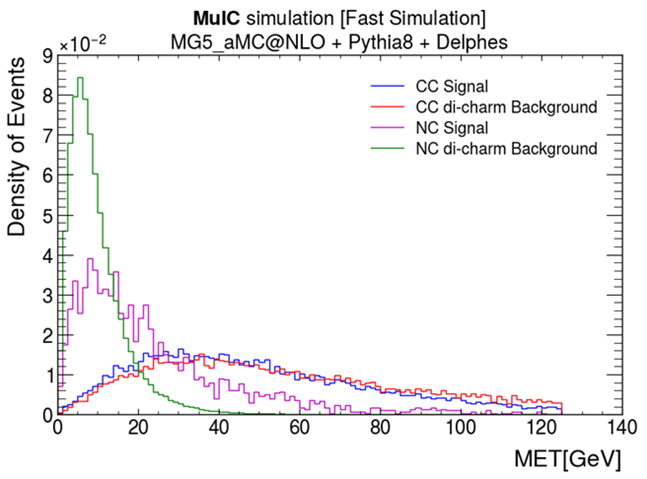}
    \caption{Comparison of MET between di-charm background and \Hcc\ signal in both CC DIS and NC DIS production modes.}
    \label{fig:Neutral Current Background and Higgs}
\end{figure}

Our fast simulations of CC DIS signal and di-charm background suggests there is little benefit in separating these two populations based solely on the shape of the \met\ distribution. However, the distinction between CC DIS signal and NC DIS di-charm background is far more significant. NC DIS events maintain the beam muon and produce little or no real \met, except from final-state physics involving one or more low-energy neutrinos or from mis-reconstruction of the event. 
The \met\ in these two populations is compared in Fig.~\ref{fig:Neutral Current Background and Higgs}. As expected, \met\ peaks more sharply toward zero in the NC DIS di-charm background, while it extends to typically much larger values in CC DIS signal simulation.

We hypothesized that there might be additional benefits in looking at the angular distribution of the MET relative to the di-jet system that forms the Higgs candidate. We recognize that in a collision where only the Higgs and the recoiling neutrino are produced, the \met\ is entirely defined as being opposite the sum of the jet particles and thus is 100\% correlated with the di-jet system. However, in collisions where part or all of the beam remnant is present, or where additional radiation or other particles appear in the detector, the correlation will be weakened. Naively, we anticipated that in CC DIS Higgs signal events we should expect to see a strong peak in the azimuthal angle difference between the \met\ and the Higgs candidate around $\pm \pi$. In NC DIS di-charm events, where there is no real significant source of \met\ and the possibility of a muon in the detector, we anticipated that there would be no or a much weaker correlation.

\begin{figure}[htbp]
    \centering
    \includegraphics[width=\linewidth]{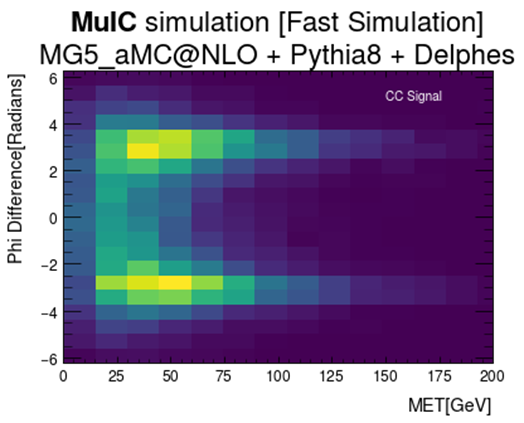}
    
    \includegraphics[width=\linewidth]{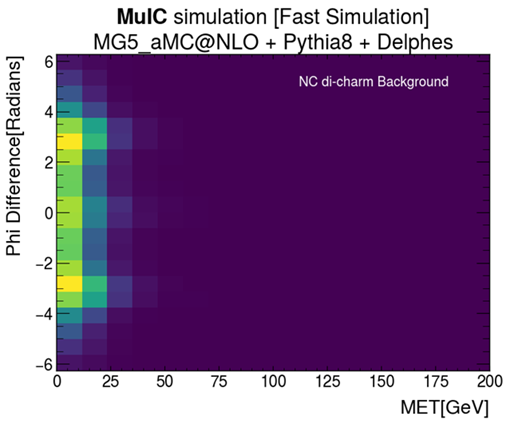}

    \caption{A comparison of $\Delta \Phi_{\mathrm{MET,jj}}$ vs \met\ for (top) CC DIS signal and (bottom) NC DIS di-charm background simulation.}
    \label{fig:met_phi_heatmap}
\end{figure}

We show the results of a basic study of this in Fig.~\ref{fig:met_phi_heatmap}.
As predicted, the difference in the missing energy angle for CC DIS signal peaked at $\pm \pi$, while the NC DIS di-charm background was somewhat more diffuse. Nevertheless, we still observe strong $\pm \pi$ values for this angular difference in the transverse plane. We do not exploit this possible distinction in angular correlations for the remainder of this note, but we point out that such a variable (with additional refinement) may be of future use in an analysis that targets extreme separation between CC DIS signal and NC DIS di-charm background.

\section{A Prototype Data Analysis for \Hcc}
\label{sec:analysis}

We present here the outcome of a very basic event selection and candidate refinement. This is meant to provide a conservative view of the potential for selecting \Hcc\ candidates, with emphasis on CC DIS production (the leading cross-section). We apply a basic optimization procedure to set criteria at various stages of the analysis, employing the method of varying a criterion to maximize the following figure of merit~\cite{Cowan:2010js}
\begin{equation}
    \sqrt{2\left((s+b)\cdot\ln(1+s/b) - s \right)}
\end{equation}
where $s$ ($b$) is the number of events in the signal (background) class. This quantity is the asymptotic formula for significance in the limit of large backgrounds, which is a reasonable assumption in this study. The class definitions vary depending on the goal of the selection. We conservatively round any criteria determined from this procedure to avoid over-interpretation of the optimization results on current simulation approaches and statistical limitations (e.g., $66\gev$ becomes $70\gev$ on an energy criterion).

\begin{figure}[htbp]
    \centering
    \includegraphics[width=\linewidth]{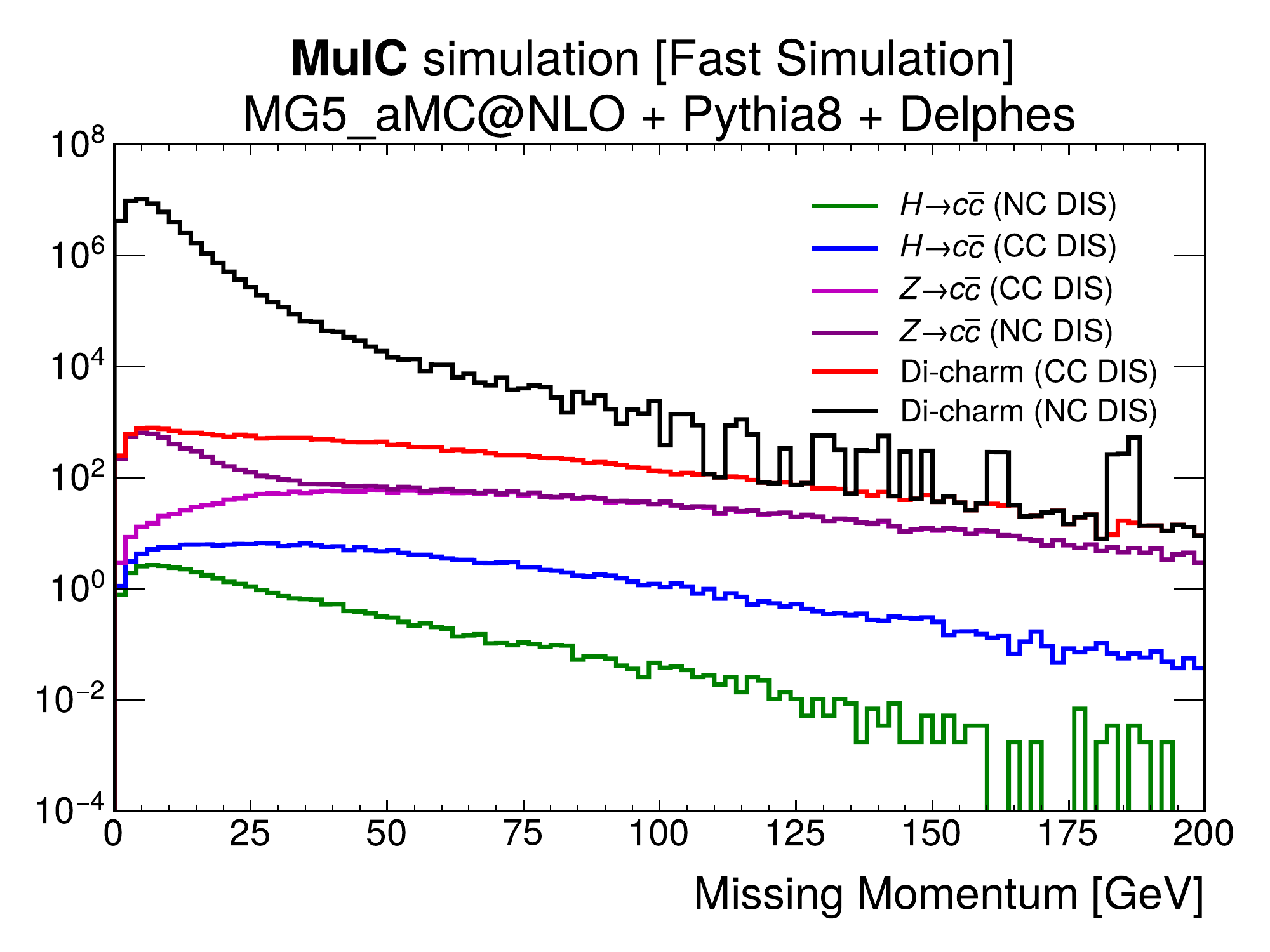}
    \caption{\met\ with no other selection criteria applied for a target integrated luminosity of $100\invfb$.}
    \label{fig:met_presel}
\end{figure}

\begin{figure}[htbp]
    \centering
    \includegraphics[width=\linewidth]{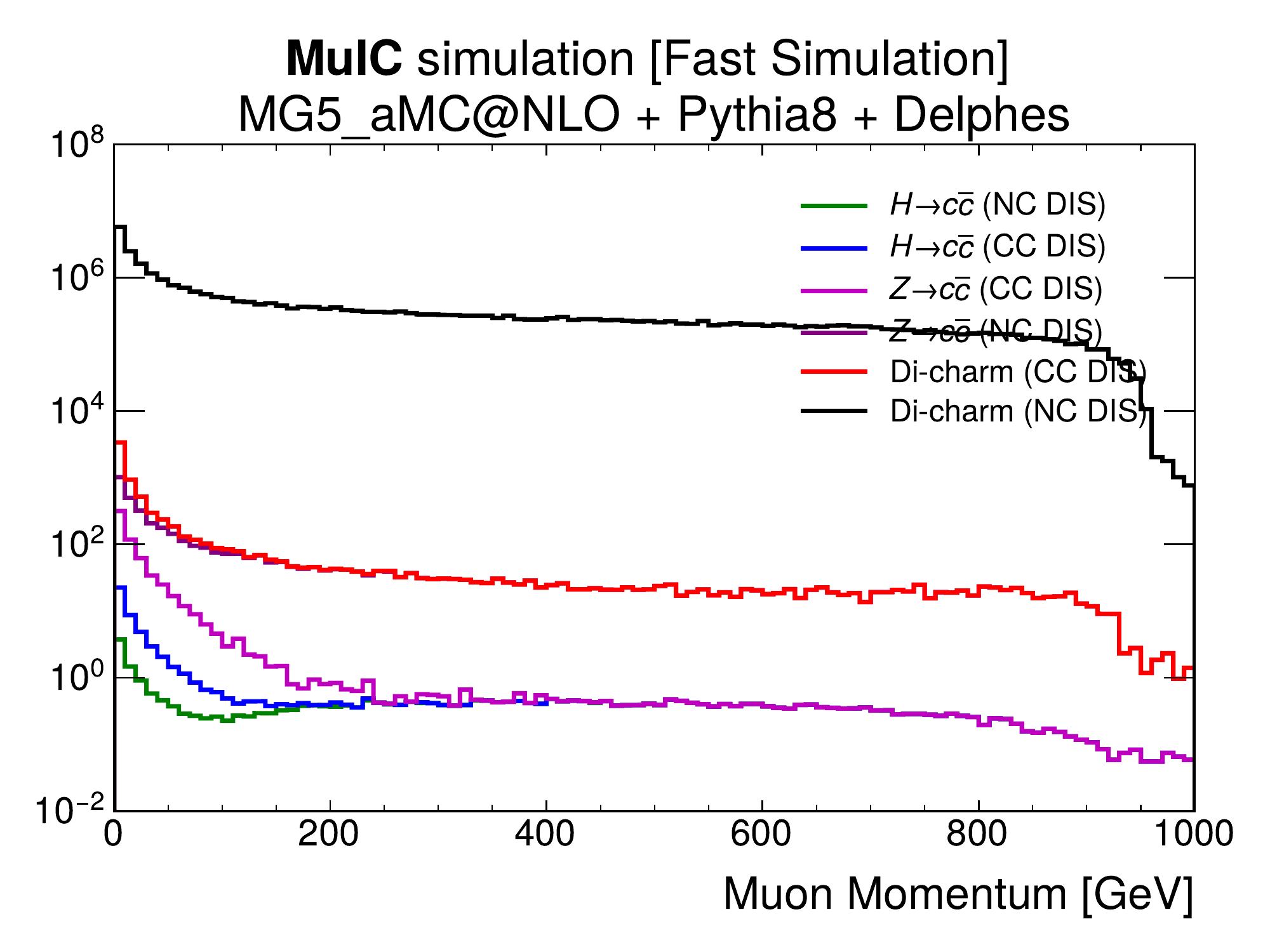}
    \caption{Muon momentum with no other selection criteria applied for a target integrated luminosity of $100\invfb$.}
    \label{fig:mup3_presel}
\end{figure}

The most basic event characteristics are that the muon-proton collisions, through the CC DIS process, will produce a significant and real (e.g. due to at least one high-momentum neutrino) quantity of missing momentum (Fig.~\ref{fig:met_presel}). In addition, the event should not contain evidence of a high-momentum muon at low angle in the backward direction; that signature is a hallmark of the NC DIS process (Fig.~\ref{fig:mup3_presel}). Given the lower cross-section for Higgs production in NC DIS and the substantially increased rate of production of non-resonant di-charm and \Zcc\ in NC DIS, our goal is to minimize contamination from NC DIS events. We define the signal class as containing luminosity-weighted CC DIS $\Hcc$ events and the background class to contain NC DIS $\Zcc$ and inclusive di-charm events. We find that $\met>50~\gev$ and rejecting events with a backward (bwd) muon satisfying $p^{bwd}_{\mu} > 60\gev$ optimizes our criteria. The optimizations are not performed accounting for correlations in the two variables and are only one-sided in each variable (a minimum in \met\ and maximum in $p_{\mu}$). We refer to the selection on MET and the muon veto as the "Event Tag."

\onecolumngrid

\begin{table}[tbp]
    \centering
    \begin{tabular}{l|c|c|c|c|c|c}
    & & \multicolumn{5}{c}{Yield ($100~\invfb$)/Total Efficiency [\%]/Relative Efficiency [\%]} \\
    
     Selection  &  Criteria & \Hcc\ & \Zcc\ (CC DIS) & \Zcc\ (NC DIS) & 
     Incl. \ccbar (CC DIS) & 
     Incl. \ccbar (NC DIS) \\

    \hline

     No Selection & 
     None & 
     188/-/- & 
     2800/-/-& 
     4500/-/-& 
     15000/-/- & 
     $5\times 10^7$/-/- \\

     Event Tag & 
    \begin{minipage}[t]{0.15\textwidth}
     $\met>50~\gev$
     \\
     $p_{\mu}^{bwd}<60~\gev$ 
    \end{minipage} &
     79/42/42 & 
     1,900/68/68 & 
     25/0.6/0.6 & 
     7,000/45/45& 
     51,000/0.1/0.1\\

    Jets &
    \begin{minipage}[t]{0.15\textwidth}
       $|\eta_{j}|<4.0$
        \\
        $p_T^j > 20~\gev$
    \end{minipage} &
     76/40/96 & 
     1,800/65/96 & 
     23/0.5/93 & 
     3,700/24/53& 
     46,000/0.09/91 \\

     Charm Dijet & 
     \begin{minipage}[t]{0.15\textwidth}
        $\eta_{j} < 0$
        \\
        2 Charm Tagged
     \end{minipage}& 
     4/2/5 &
     96/3.5/5 & 
     0.9/0.02/4 & 
     56/0.4/2 & 
     1500/0.003/3\\

     \hline
    \end{tabular}
    \caption{Selection stages, criteria, and efficiency on various samples. For each sample, and for each selection category, we report the number of events (scaled to 100\invfb\ of collision data), the cumulative efficiency (in percent) of the selection, and the relative efficiency (in percent) compared to the previous selection category. Given the limited statistics available in the NC DIS dicharm background, we note that only 6 unweighted events survive all selection criteria; the resulting prediction for that sample therefore has a 41\% MC simulation statistical uncertainty.}
    \label{tab:selection}
\end{table}

\twocolumngrid

Candidate events should then contain at least two jets within the fiducial tracking volume and with a sufficient minimum \pt\ to be compatible with the decay of a heavy resonance. This is referred to as the "Jets Selection." Among those jets, there should be at least two charm-tagged jets. These jets should be located primarily in the backward direction. Constraints that emphasize the selection of events consistent with these expectations are referred to as "Charm Dijet Selection."

Finally, we expect the dijet system to form an invariant mass consistent with the Higgs particle. Constraints on the event that emphasize specifically the Higgs signature are referred to as the "Higgs Selection."

The selection criteria and their efficiency on the three main samples --- \Hcc, \Zcc, and di-charm --- are shown in Table~\ref{tab:selection}. The samples are weighted to a target luminosity of $100~\invfb$. We note that the NC DIS di-charm background was the most difficult to simulate in sufficient quantity for this basic study. Whether using the more inclusive sample or the sample where $m_{cc}>50\gev$ at generator-level, we obtain almost the same number of raw events passing all selection criteria: about 6. This means any weighted prediction using these samples have about a 40\% MC simulation statistical uncertainty and can only be used to guide thinking at this stage.

We also simulated NC DIS \Hcc\ production (not shown in Table~\ref{tab:selection}), but as the analysis was designed to eliminate NC DIS production this resulted in a negligible contribution to the overall signal yield (about 1\% of the \Hcc\ yield). We note that where we expect consistent performance in selection criteria between samples, we do observe it, giving confidence to the underlying simulation models. For example, we expect the event tag criteria to substantially eliminate NC DIS; they do, maintaining CC DIS in any sample at a high level of efficiency while suppressing NC DIS-based samples. We expect that charm identification will perform with about the same efficiency on the $H$ and $Z$ simulation samples, since each contains high-momentum real charm-initiated jets. Indeed, we see about the same performance in both samples (a 5\% di-charm efficiency).

Based on our simulations, the most promising target for a future MuIC \Hcc\ search would be to emphasize the CC DIS production mechanism. This conclusion is based solely on the challenge posed in rejection NC DIS di-charm background from this analysis. Missing momentum and muon identification will be essential ingredients in this effort; the performance on these two physical observables, in turn, is defined by the quality of the calorimeter (and tracking, for particle flow) as well as the coverage of a backward muon system dedicated to beam-muon tagging. Our approach has utilized a calorimeter of reduced granularity compared to alternative models emphasized for the EIC detectors and is therefore expected to provide a much more conservative performance on MET than we think is achievable using the kind of calorimeter designs anticipated for EIC experiments.

Proposed EIC detectors do not emphasize muon reconstruction with low-angle coverage, so the deployment of such a dedicated muon system will be necessary for the MuIC. We see from Table~\ref{tab:selection} that the NC DIS di-charm background is of particular concern owning to its substantial cross-section in the jet fiducial region of interest. Reducing this background will rely heavily on efficient event tagging with a strong separation between NC and CC DIS candidate events.

We note that even at early stages of the selection the limited statistics of the di-charm sample are apparent. It was not possible for this basic study to simulate the target luminosity's worth of inclusive di-charm events, forcing us to apply large weights to these events. The effect of limited statistics becomes greater at increasingly restrictive selection. In our figures, for any distribution beyond the event tag stage we we utilize the high-mass di-charm simulation. Even here, for the present study we have limited statistics compared to the target luminosity.

\begin{figure}[htbp]
    \centering
    \includegraphics[width=\linewidth]{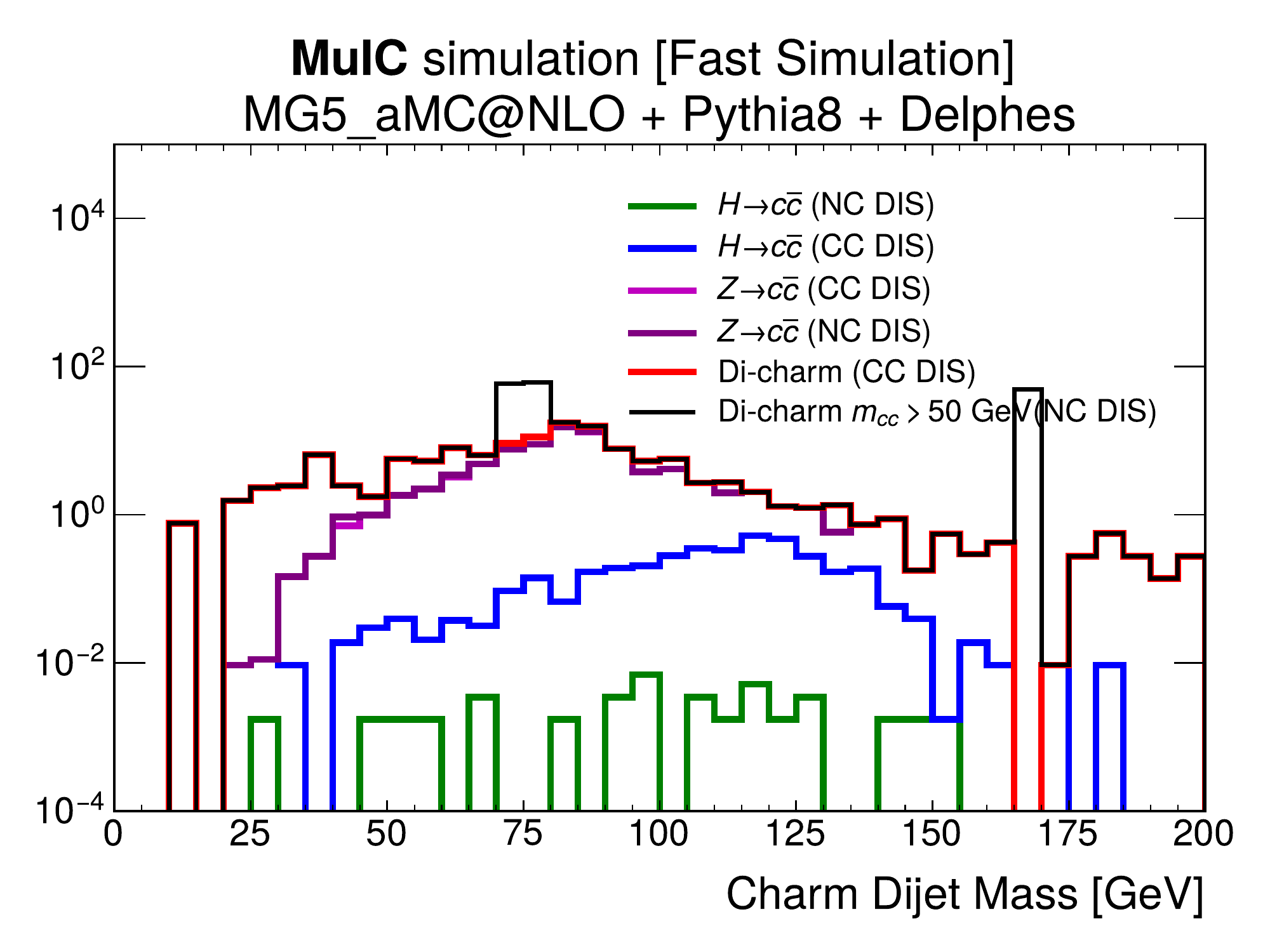}
    \caption{Charm-tagged di-jet mass with all selection criteria applied for a target integrated luminosity of $100\invfb$. The NC DIS di-charm background is not reliable owning to the low statistics of the population surviving all selections.}
    \label{fig:hcc_mass}
\end{figure}

We show in Fig.~\ref{fig:hcc_mass} a distribution of the charm-tagged di-jet mass. We recognize first and foremost that it is difficult to draw conclusions about the NC DIS di-charm background; nevertheless, the earlier stages of event selection suggest that even at this most restricted stage one should expect this population to dominate in the current prototype analysis. Additional resources will be needed to simulate sufficient samples of this background and additional approaches, such as more information and/or a correlated multivariate analysis, will be needed to improve selection of signal and rejection of this overwhelming background.

Beyond the NC DIS di-charm background it is clear that work will also be needed to improve the ability to separate $Z$-boson-induced backgrounds from the \Hcc\ signal. Separation of these topologies at the LHC has made substantial progress in the past decade, and no doubt a refined approached can be brought to the question of the viability of a future MuIC search for \Hcc. In many ways, this measurement at the MuIC may be as challenging as was the original measurement of \Hbb\ at the LHC. The accomplishment of that task within a decade of starting the LHC program provides lessons for this future-looking issue.

\section{Conclusions}

We have studied the production of the \Hcc\ final state at a future MuIC facility using a conservative baseline detector inspired by earlier work for the Snomass community planning exercise. We have assumed rather modest (27\%) charm jet-tagging efficiency, which is generally considered to be the minimum level of performance a modern algorithm can manage. We have assumed a more coarse calorimeter than those generally planned for EIC experiments, but in addition to baseline EIC-like detector configurations we have added a low-angle LHCb-like muon system to a fast simulation of an experiment.

We identify NC DIS non-resonant di-charm background as the leading challenge in any MuIC search for \Hcc. The second-leading background is expected to arise from \Zcc\ production. We have not explored in this note the reality that the $b \to c$ jet mis-identification rate, typically at the level of a 10\% or so, will mean an additional background both from \Zbb\ and, more importantly, \Hbb\ events. Given existing approaches to simultaneous fits of di-jet samples for both $b$ and $c$ final states, we have confidence that should other backgrounds be reduced this one can be resolved through the use of a simultaneous fit of a Higgs candidate sample for both final states, including mis-identification rates assessed on independent samples.

We comment on an element that is missing from this study and that would necessarily be a target for future work. The charm-tagging mis-identification rate for light jets in this study was 1.6\%. Naively, the light di-jet mis-identification rate would be than about $2.6 \times 10^{-4}$. However, the light di-jet production cross-section from NC DIS or CC DIS will be much larger than the cross sections studied for background processes in this note. We expect a significant challenge from light di-jet background in a future \Hcc\ analysis, one that will also be especially difficult to simulate in sufficient quantity and that might benefit from entirely data-driven approaches in a real data analysis. We leave this as a comment for future experimental physicists.

We conclude from this preliminary work that any future MuIC experiment must have good calorimeter granularity and readout capability, good tracking coverage for the purpose of both particle-flow reconstruction and jet acceptance (especially for charm-jet tagging performance at lower angles), and muon identification capability at low angle. Nevertheless, these are necessary but not sufficient ingredients for ultimately selecting \Hcc\ candidates in data. The challenge of background reduction, rejection, and characterization will provide daunting if such a search is undertaken. We caution the use of \Hcc\ as a motivation for a future MuIC program. However, we advocate for future consideration of this channel as part of a healthy MuIC physics analysis portfolio.
\section{Acknowledgements}

We are extremely grateful to Darin Acosta, Osvaldo Miguel Colin, and Xunwu Zuo for valuable conversations and insights about Monte Carlo simulation and underlying physics issues in the production processes for the Higgs. We are extremely grateful to Fred Olness for valuable discussions about PDFs in the study of proton structure and Higgs production. We also acknowledge the support of the Department of Physics at SMU especially for providing facilities to support the research activities of the authors. J. Choi expresses gratitude toward Highland Park High School for its encouragement of academic excellence and especially for its Science and Technology Festival, which resulted in becoming a part of this project. P. Ahluwahlia expresses gratitude toward the Harmony School of Innovation  for its encouragement of academic excellence and independent research. We are grateful to the SMU Office of Information Technology and the Center for Research Computation at SMU for access to and support on the ManeFrame II supercomputer cluster.

\newpage

\bibliographystyle{apsrev4-1} 

\bibliography{bibliography/main}

\clearpage
\appendix

\onecolumngrid

\section{\mgamcnlo\ Feynman Diagrams}
\label{sec:feynman_mg5}

This appendix contains the Feynman diagrams that are employed by \mgamcnlo\ in the calculation of the matrix element, and thus the cross-section, for Higgs production and background processes of interest to the effort described in this note. 

The following is the key for the particles in the diagrams:

\begin{itemize}
    \item Quarks and gluons are represented by their usual lower-case letters (u, d, c, s, b, t, g);
    \item Anti-quarks are denoted by the placement of a tilde next to the quark letter;
    \item Photons are denoted by $a$;
    \item Z and W bosons are denoted by lower-case versions of their letter names.
\end{itemize}

\subsection{Charged Current DIS Vector-Boson Fusion (SM5F\_NLO)}
\label{sec:ccdis_vbf_mg5}

\begin{figure}[htbp]
    \centering
    \includegraphics[width=0.4\linewidth]{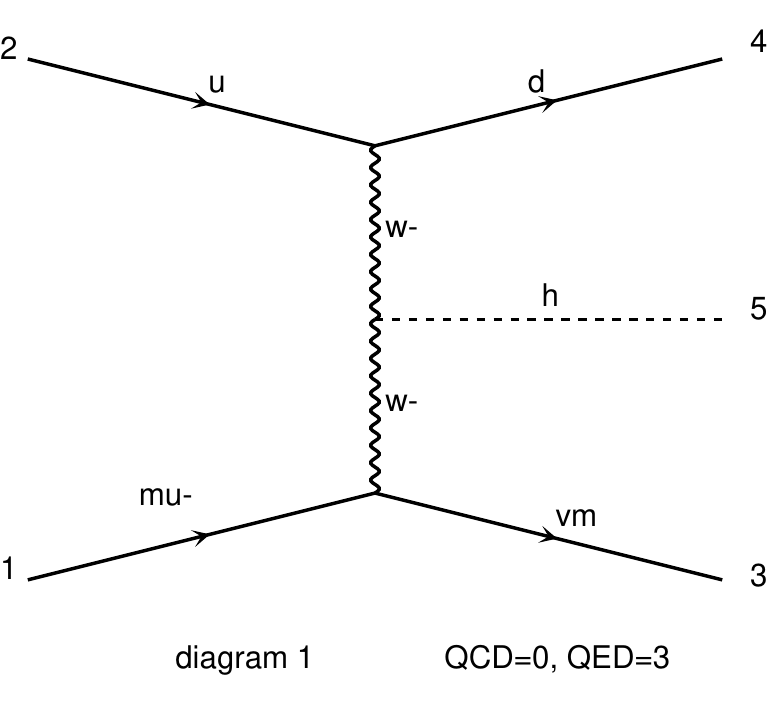}
    \hfill
    \includegraphics[width=0.4\linewidth]{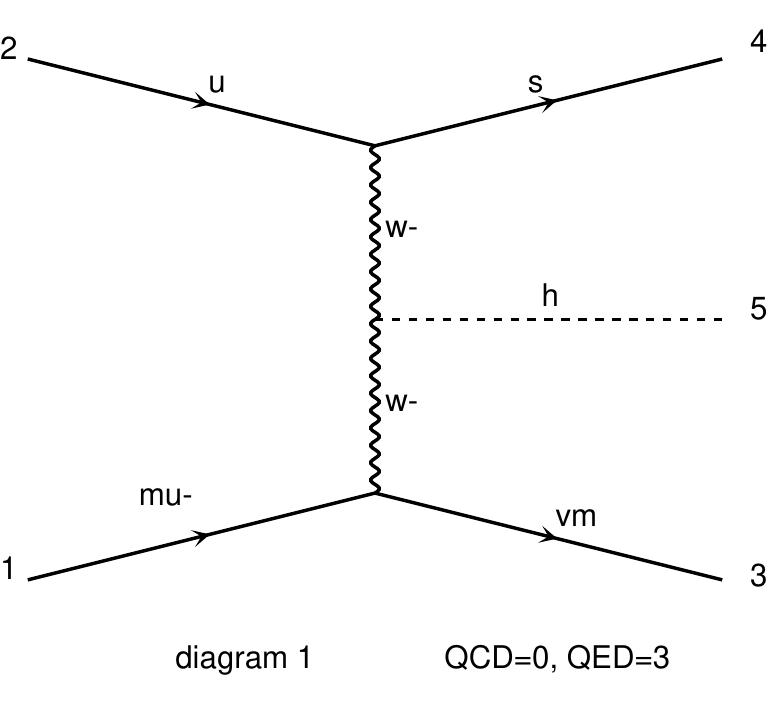}

    \includegraphics[width=0.4\linewidth]{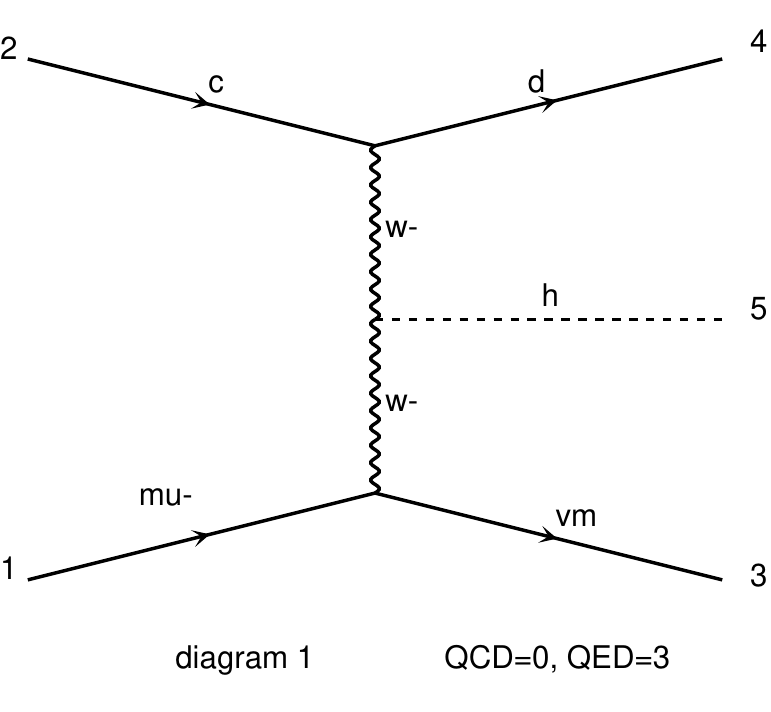}
    \hfill
    \includegraphics[width=0.4\linewidth]{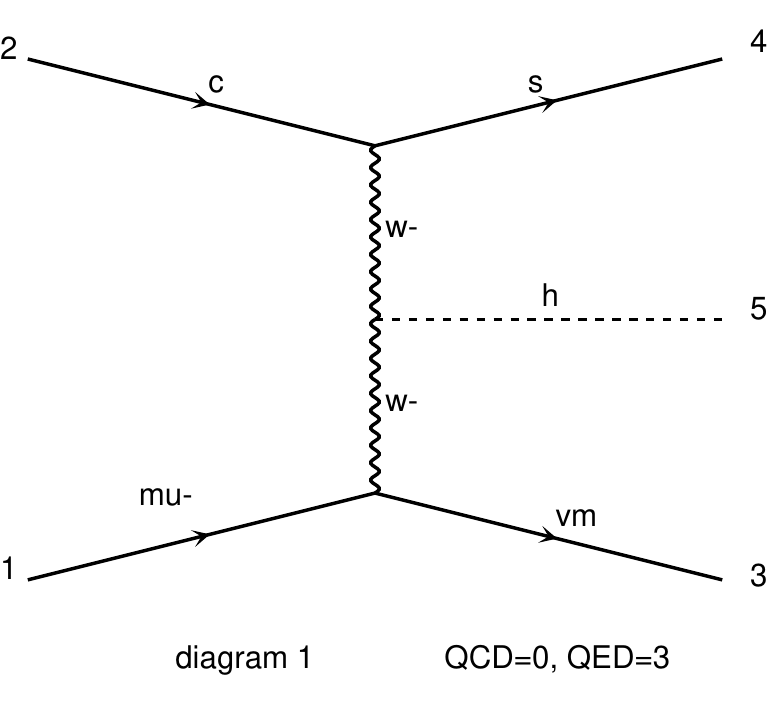}
    \caption{Diagrams used in the calculation of the CC DIS VBF process (SM5F\_NLO).}
    \label{fig:sm5f_nlo_vbf_ccdis1}
\end{figure}

\begin{figure}[htbp]
    \centering
    \includegraphics[width=0.4\linewidth]{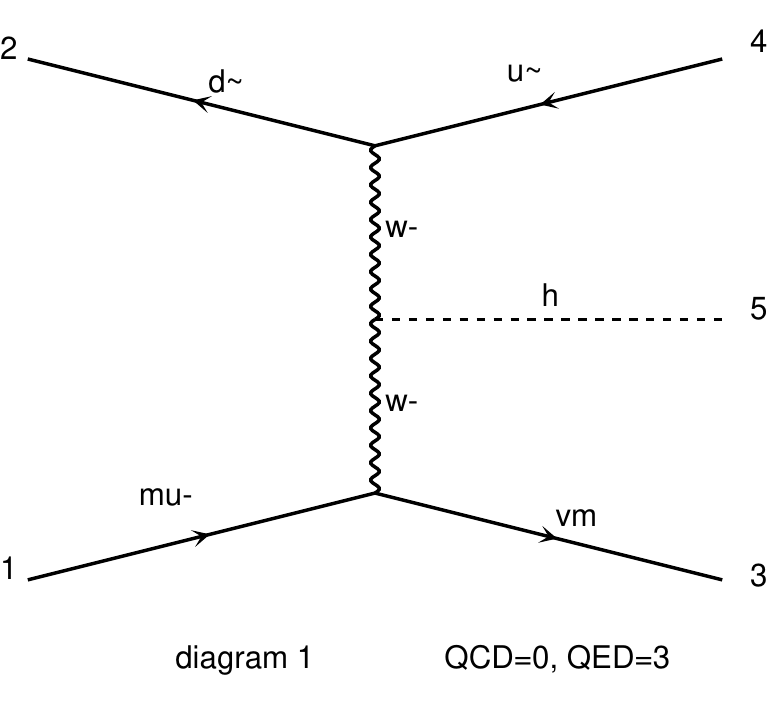}
    \hfill
    \includegraphics[width=0.4\linewidth]{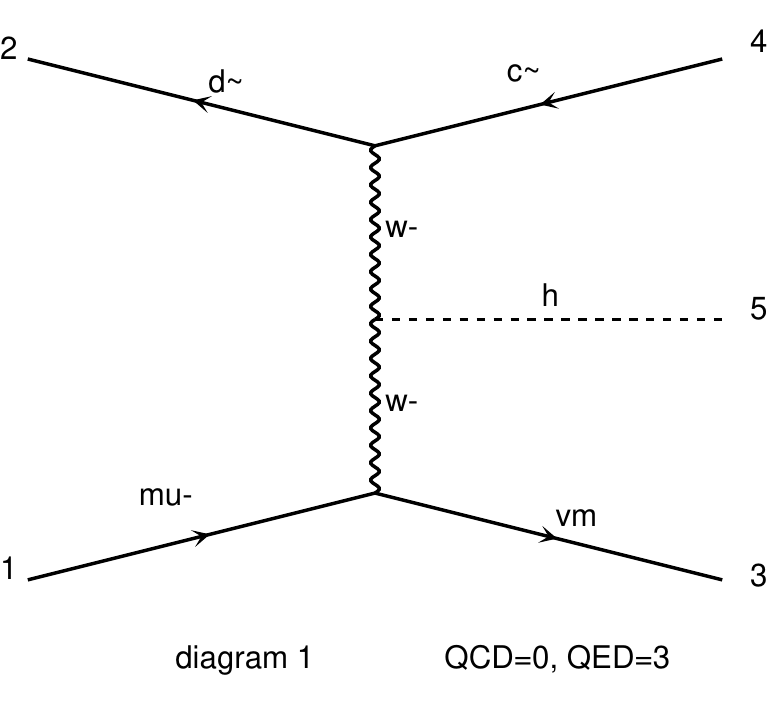}

    \includegraphics[width=0.4\linewidth]{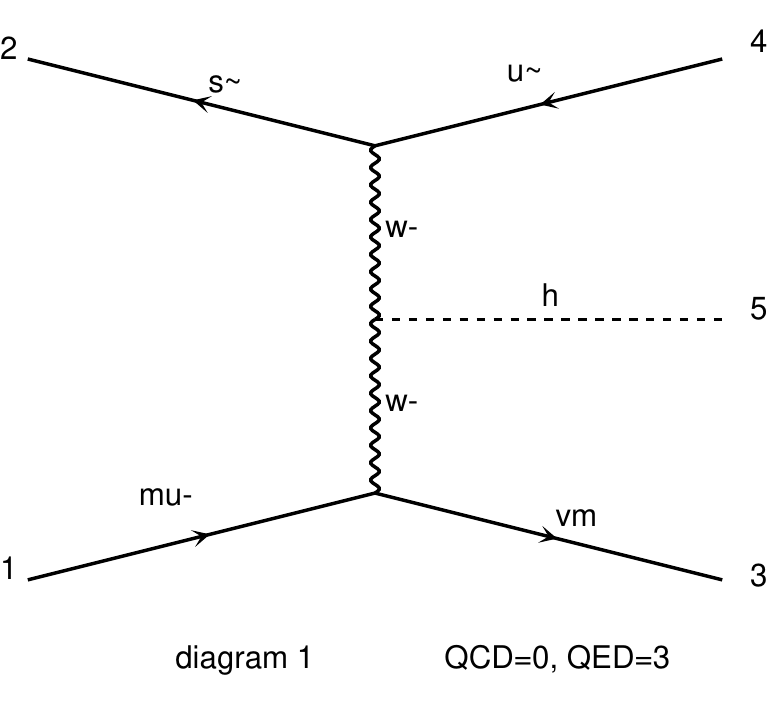}
    \hfill
    \includegraphics[width=0.4\linewidth]{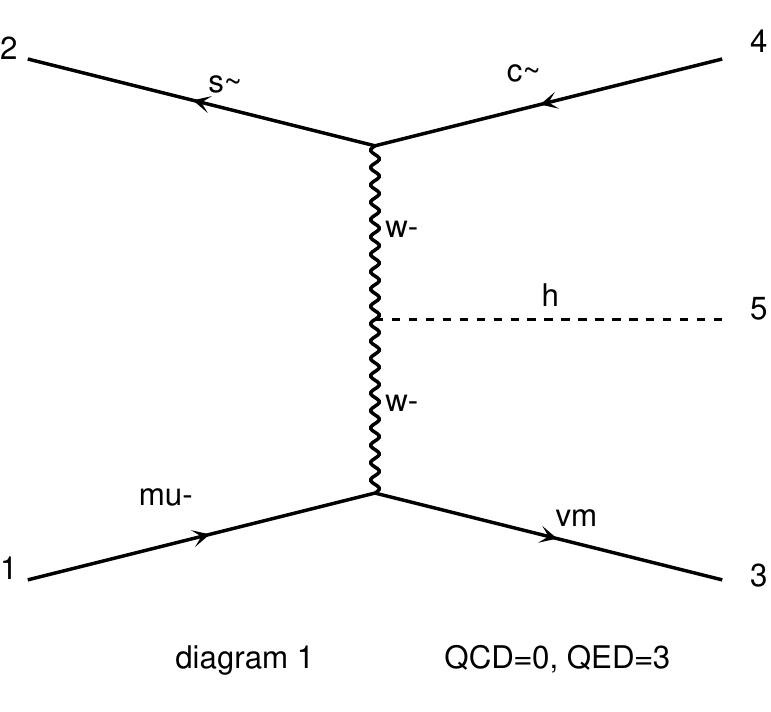}
    \caption{Diagrams used in the calculation of the CC DIS VBF process (SM5F\_NLO).}
    \label{fig:sm5f_nlo_vbf_ccdis2}
\end{figure}

\clearpage

\subsection{Neutral Current DIS Vector-Boson Fusion (SM5F\_NLO)}
\label{sec:ncdis_vbf_mg5}

\begin{figure}[htbp]
    \centering
    \includegraphics[width=0.4\linewidth]{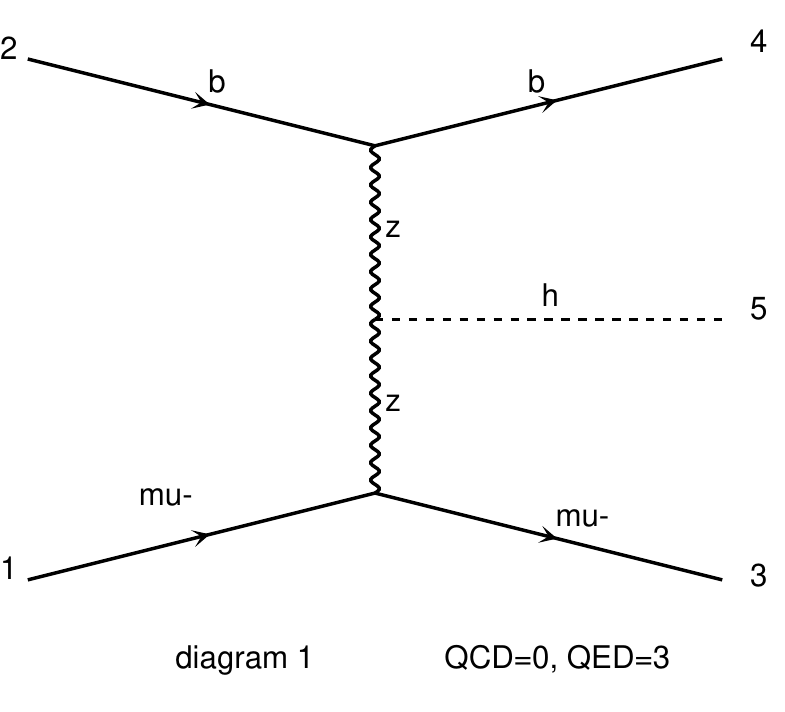}
    \hfill
    \includegraphics[width=0.4\linewidth]{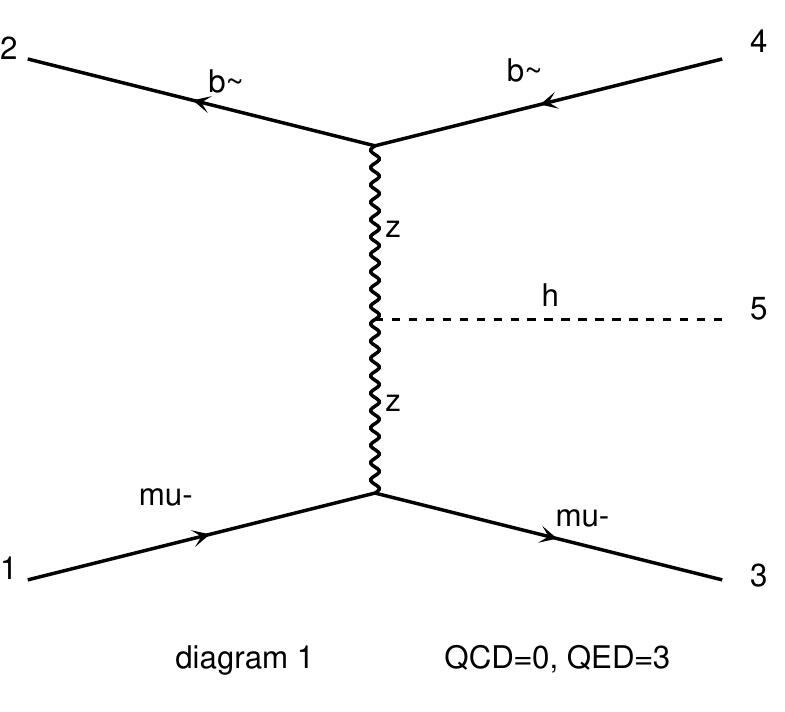}

    \includegraphics[width=0.4\linewidth]{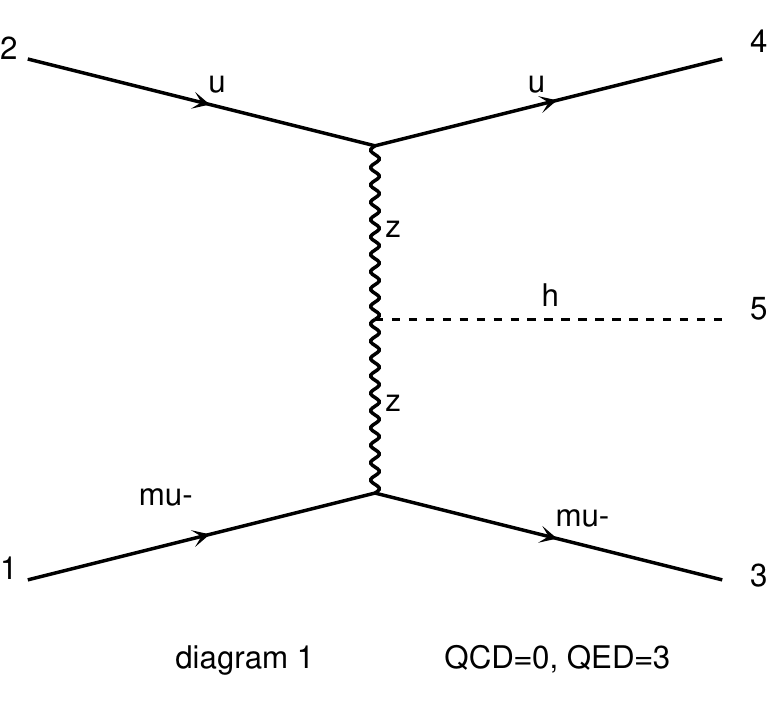}
    \hfill
    \includegraphics[width=0.4\linewidth]{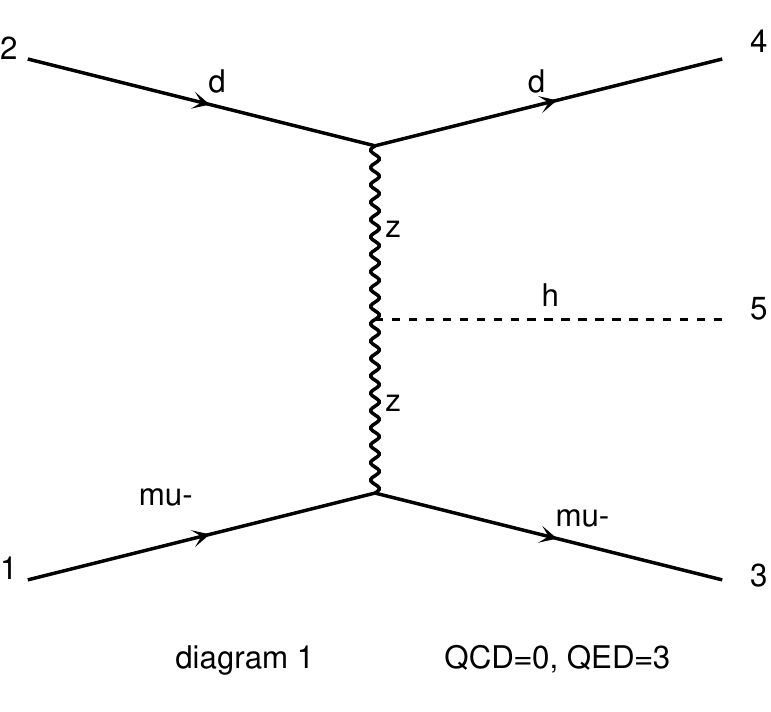}

    \caption{Diagrams used in the calculation of the NC DIS VBF process (SM5F\_NLO).}
    \label{fig:sm5f_nlo_vbf_ncdis1}
\end{figure}

\begin{figure}[htbp]
    \centering
    \includegraphics[width=0.4\linewidth]{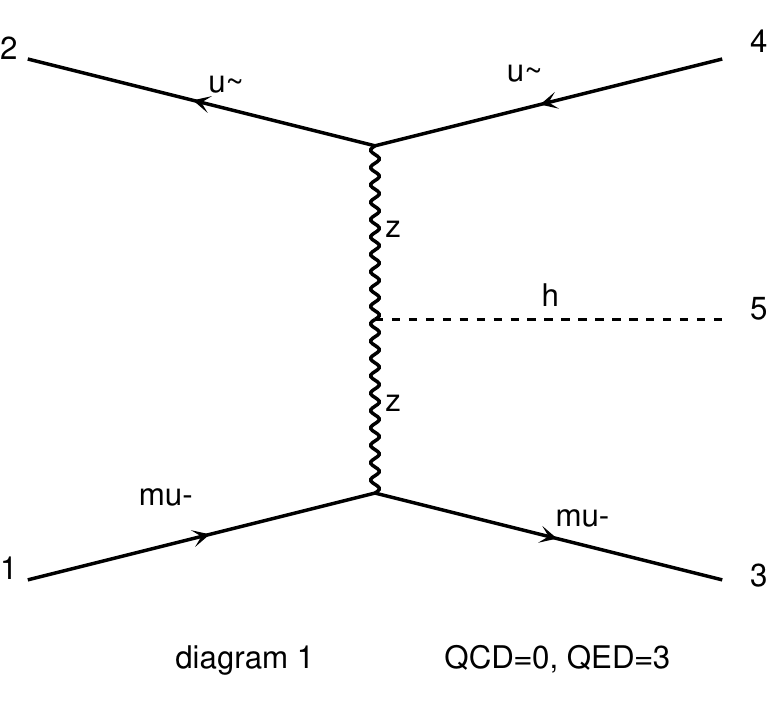}
    \hfill
    \includegraphics[width=0.4\linewidth]{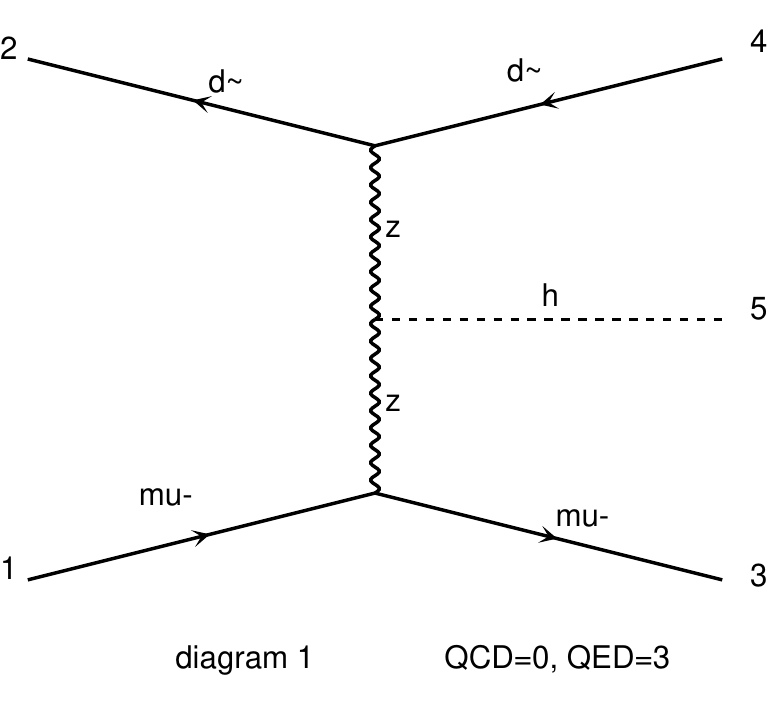}
    \caption{Diagrams used in the calculation of the CC DIS VBF process (SM5F\_NLO).}
    \label{fig:sm5f_nlo_vbf_ncdis2}
\end{figure}

\clearpage

\subsection{top-Higgs Vector-Boson Fusion (SM5F\_NLO)}
\label{sec:tH_vbf_mg5}

\begin{figure}[htbp]
    \centering
    \includegraphics[width=0.45\linewidth]{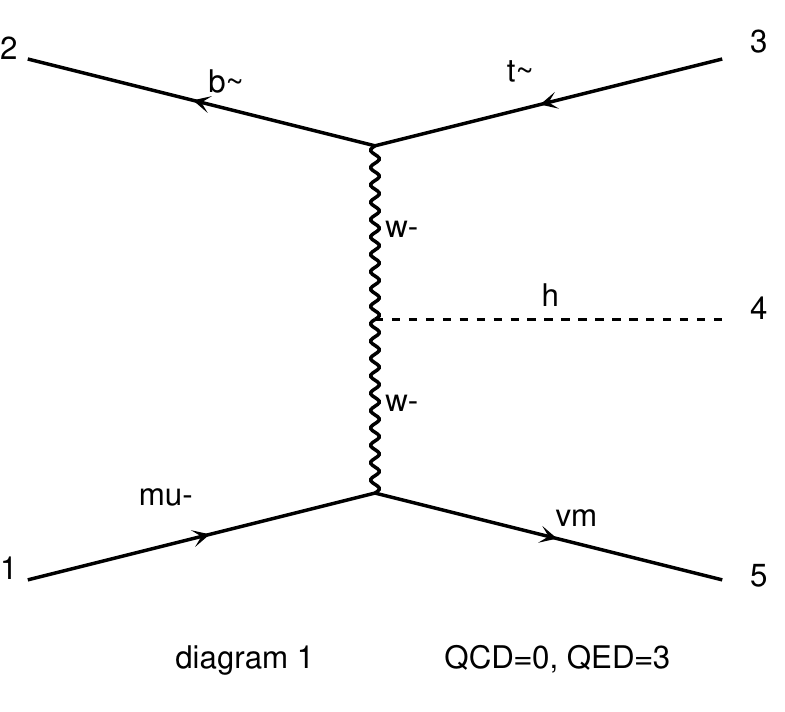}
    \caption{Diagrams used in the calculation of the top-Higgs DIS VBF process (SM5F\_NLO).}
    \label{fig:sm5f_nlo_vbf_tH}
\end{figure}

\clearpage

\subsection{Neutral-Current DIS Di-Charm Production (No Higgs Resonance)}
\label{sec:ncdis_dicharm_mg5}

\begin{figure}[htbp]
    \centering
    \includegraphics[width=0.90\linewidth]{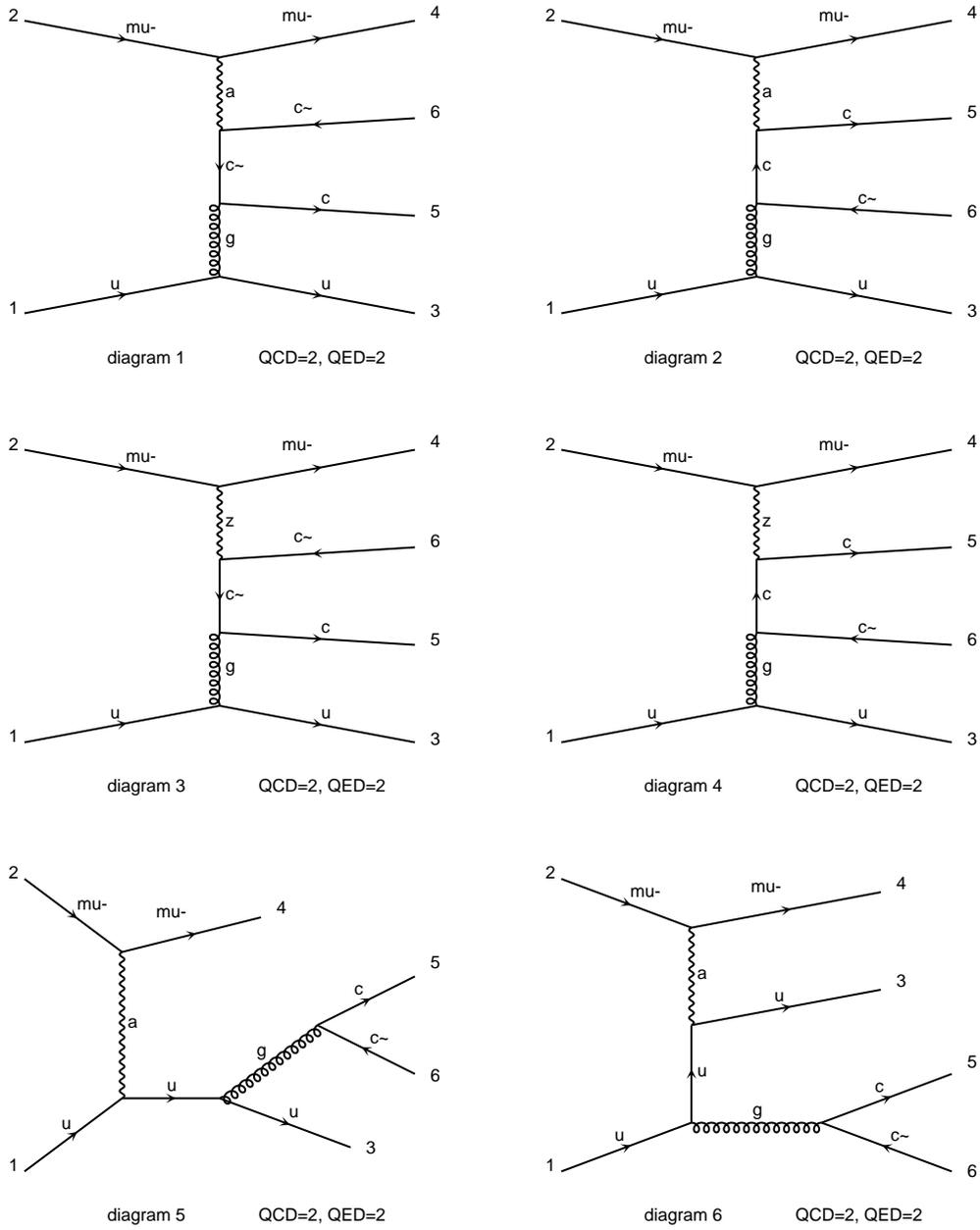}
    \caption{Diagrams used in the calculation of the NC DIS di-charm (non-resonant) process (SM5F\_NLO). The processes emphasized here are induced by up quarks in the proton.}
    \label{fig:sm5f_nlo_dicharm_ncdis1}
\end{figure}

\begin{figure}[htbp]    
    \includegraphics[width=0.90\linewidth]{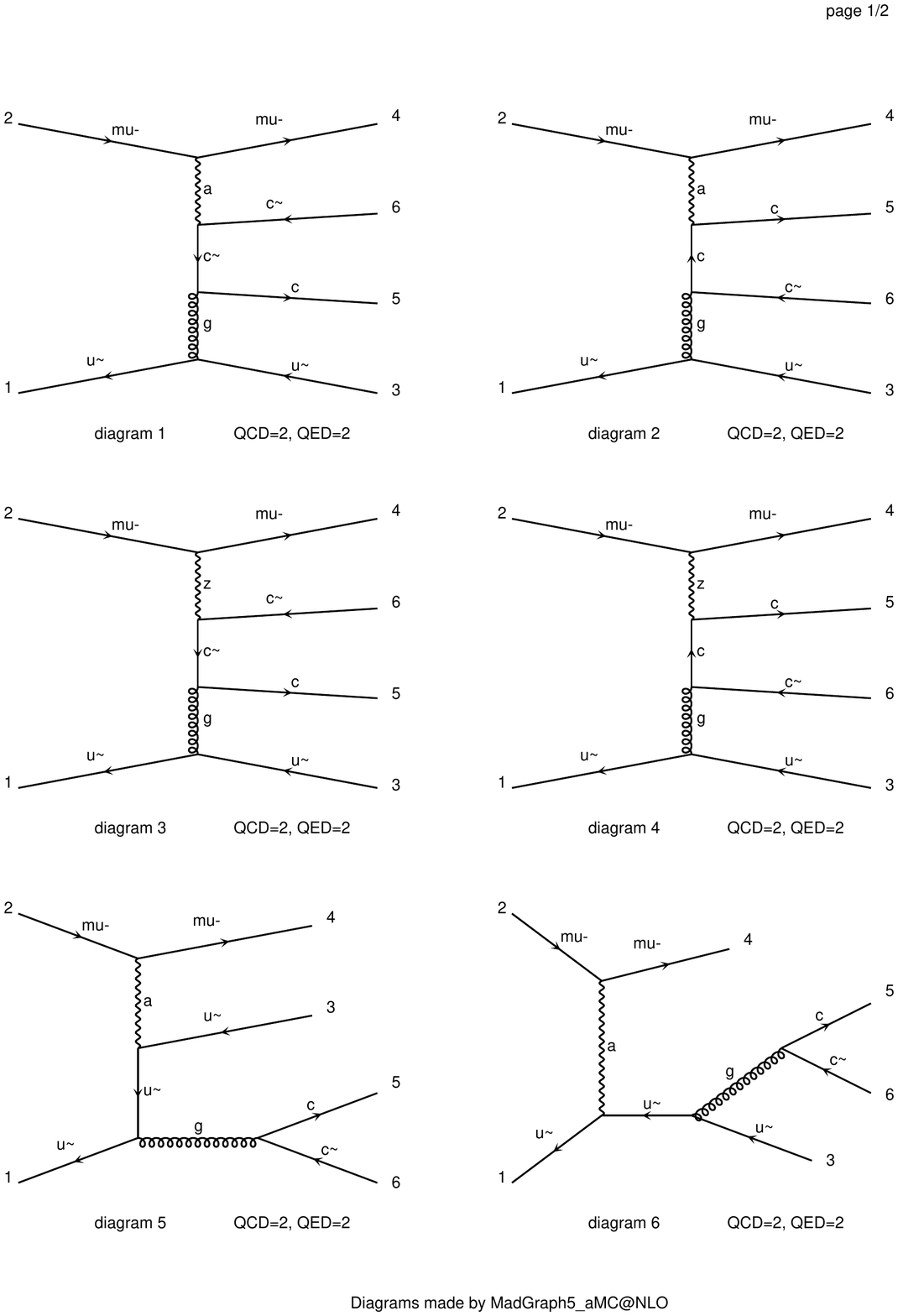}
    \caption{Diagrams used in the calculation of the NC DIS di-charm (non-resonant) process (SM5F\_NLO). The processes emphasized here are induced by up quarks in the proton.}
    \label{fig:sm5f_nlo_dicharm_ncdis2}
\end{figure}

\begin{figure}[htbp]    
    \includegraphics[width=0.90\linewidth]{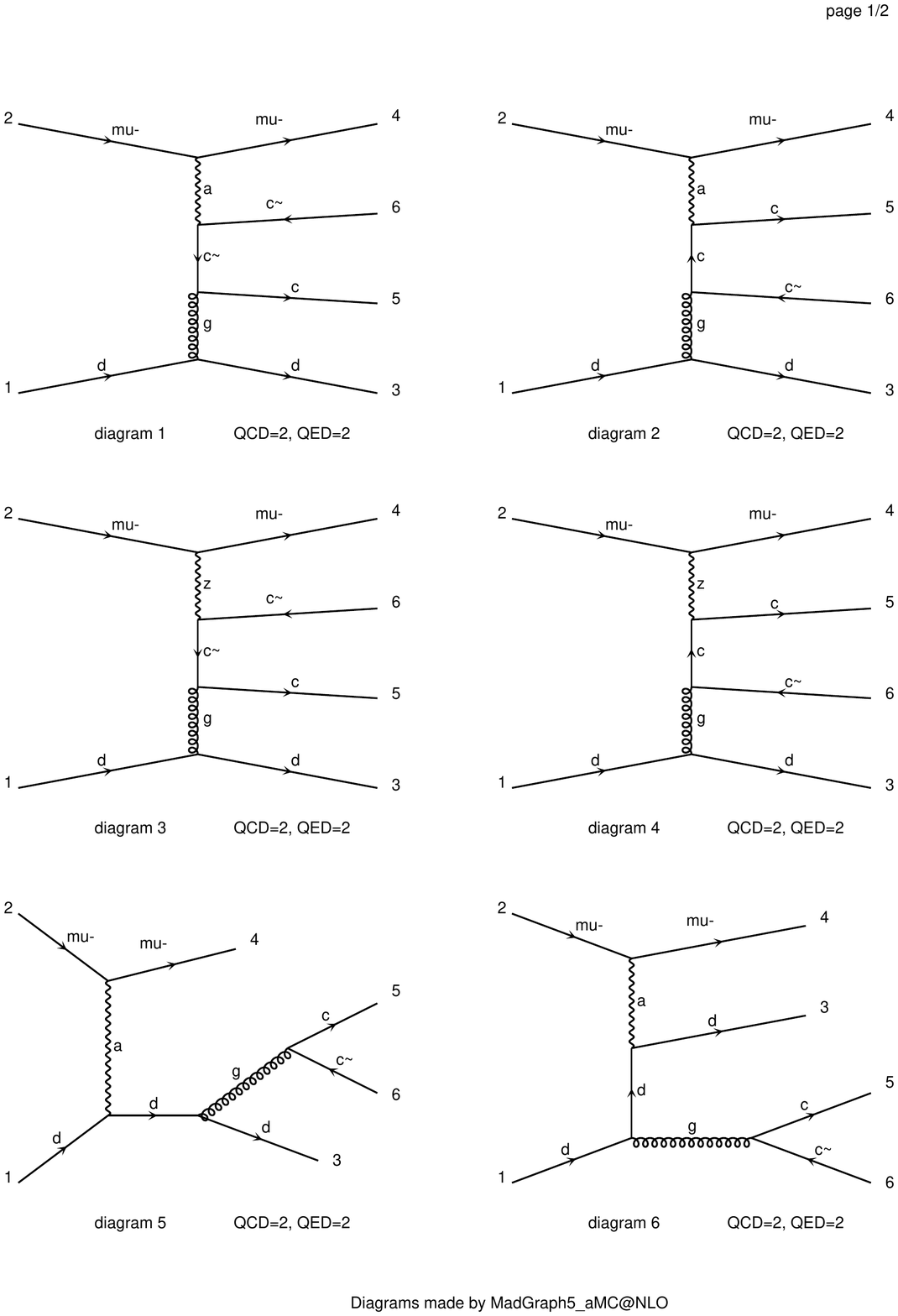}
    \caption{Diagrams used in the calculation of the NC DIS di-charm (non-resonant) process (SM5F\_NLO). The processes emphasized here are induced by down quarks in the proton.}
    \label{fig:sm5f_nlo_dicharm_ncdis3}
\end{figure}

\begin{figure}[htbp]    
    \includegraphics[width=0.90\linewidth]{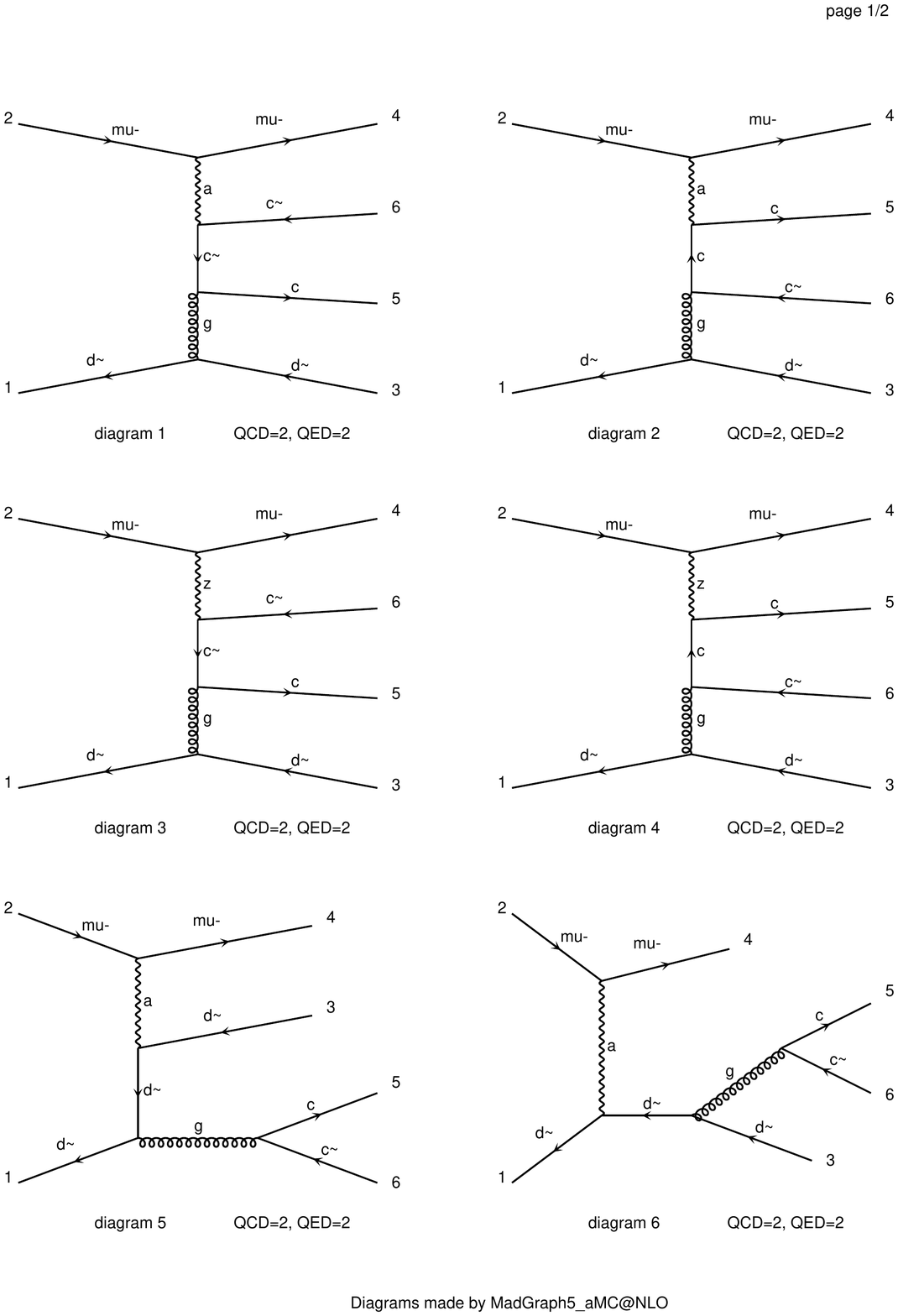}
    \caption{Diagrams used in the calculation of the NC DIS di-charm (non-resonant) process (SM5F\_NLO). The processes emphasized here are induced by down quarks in the proton.}
    \label{fig:sm5f_nlo_dicharm_ncdis4}
\end{figure}

\begin{figure}[htbp]
    \centering
    \includegraphics[width=0.90\linewidth]{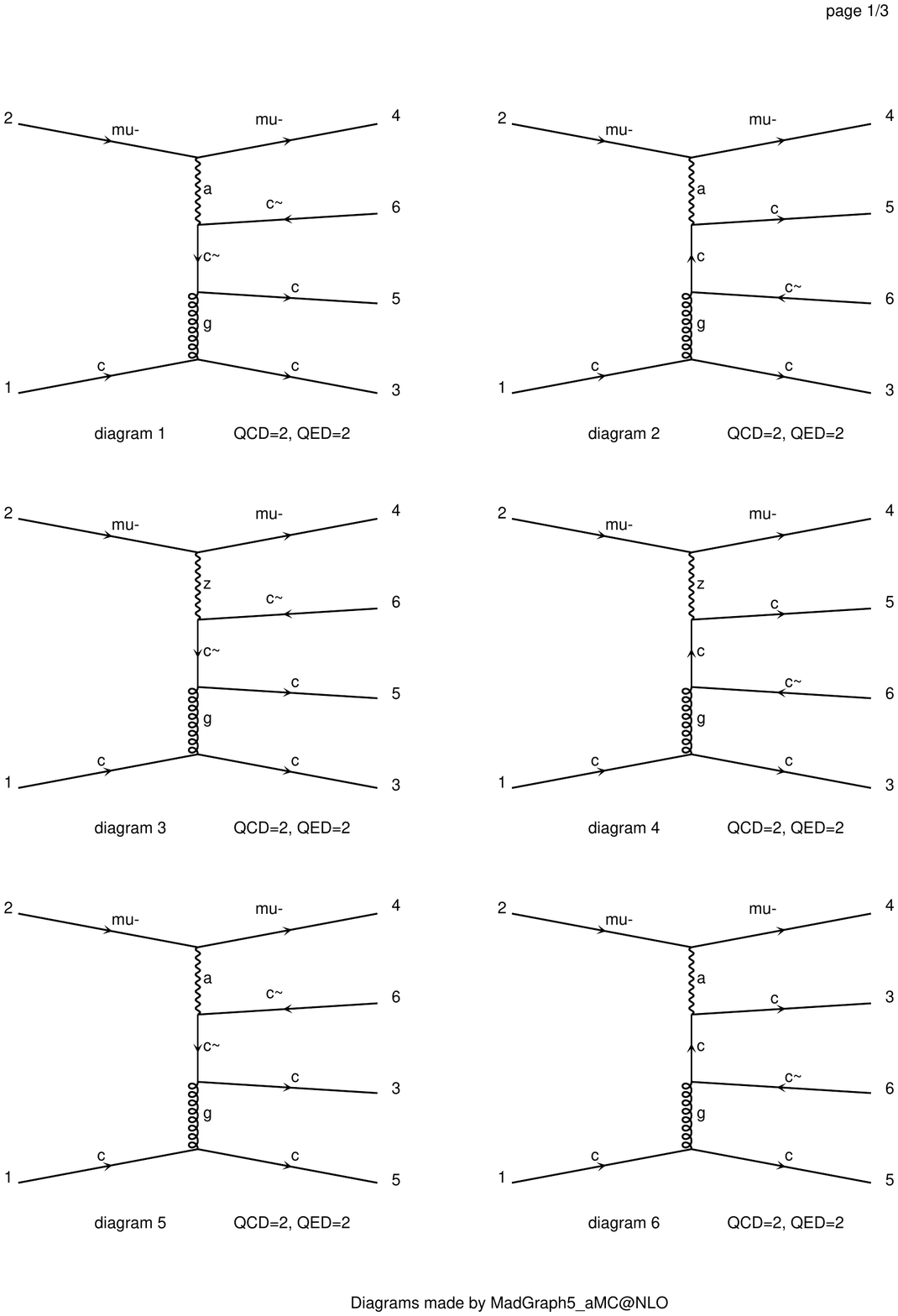}
    \caption{Diagrams used in the calculation of the NC DIS di-charm (non-resonant) process (SM5F\_NLO). The processes emphasized here are induced by charm quarks in the proton.}
    \label{fig:sm5f_nlo_dicharm_ncdis5}
\end{figure}

\begin{figure}[htbp]    
    \includegraphics[width=0.90\linewidth]{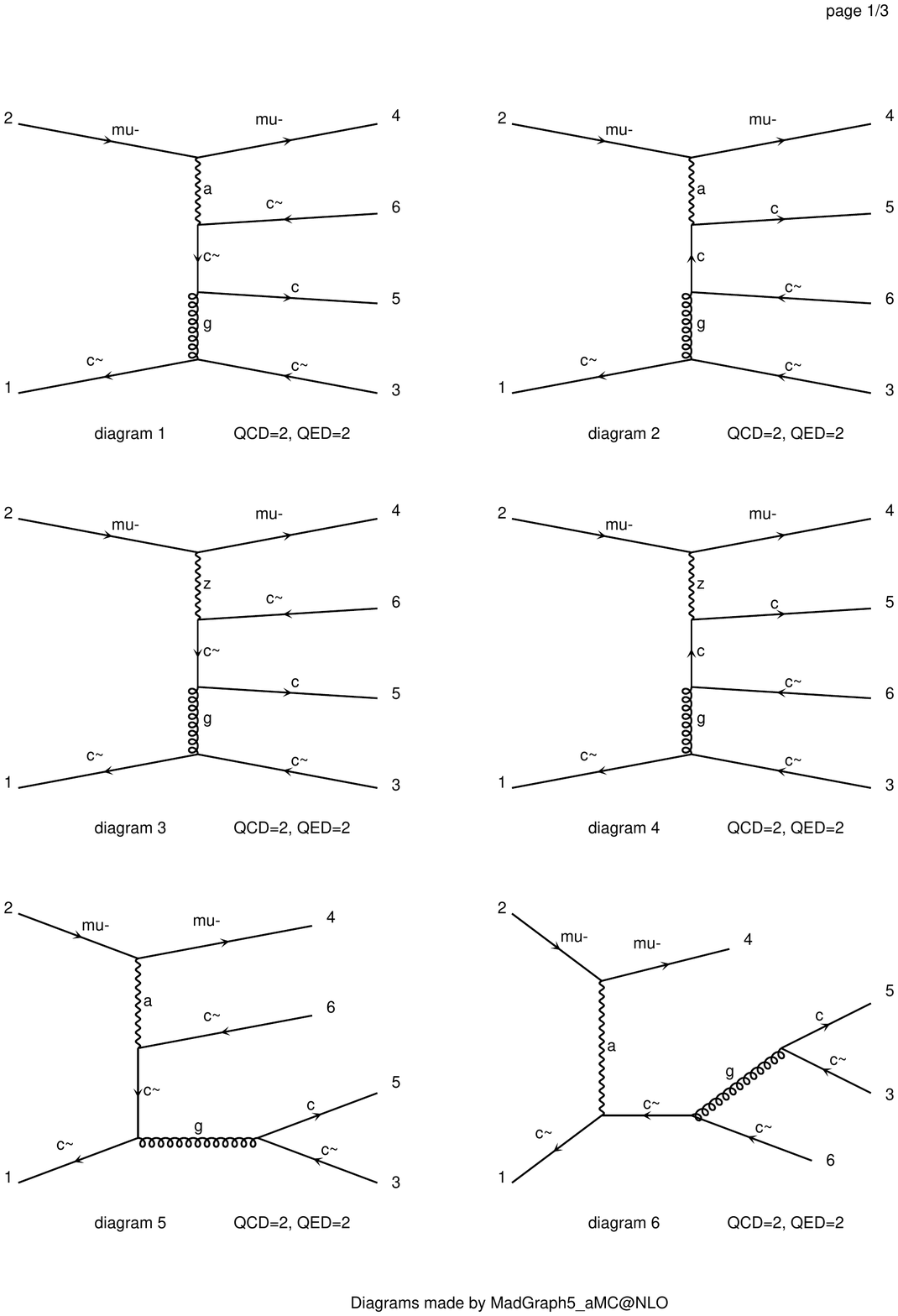}
    \caption{Diagrams used in the calculation of the NC DIS di-charm (non-resonant) process (SM5F\_NLO). The processes emphasized here are induced by charm quarks in the proton.}
    \label{fig:sm5f_nlo_dicharm_ncdis6}
\end{figure}

\begin{figure}[htbp]    
    \includegraphics[width=0.90\linewidth]{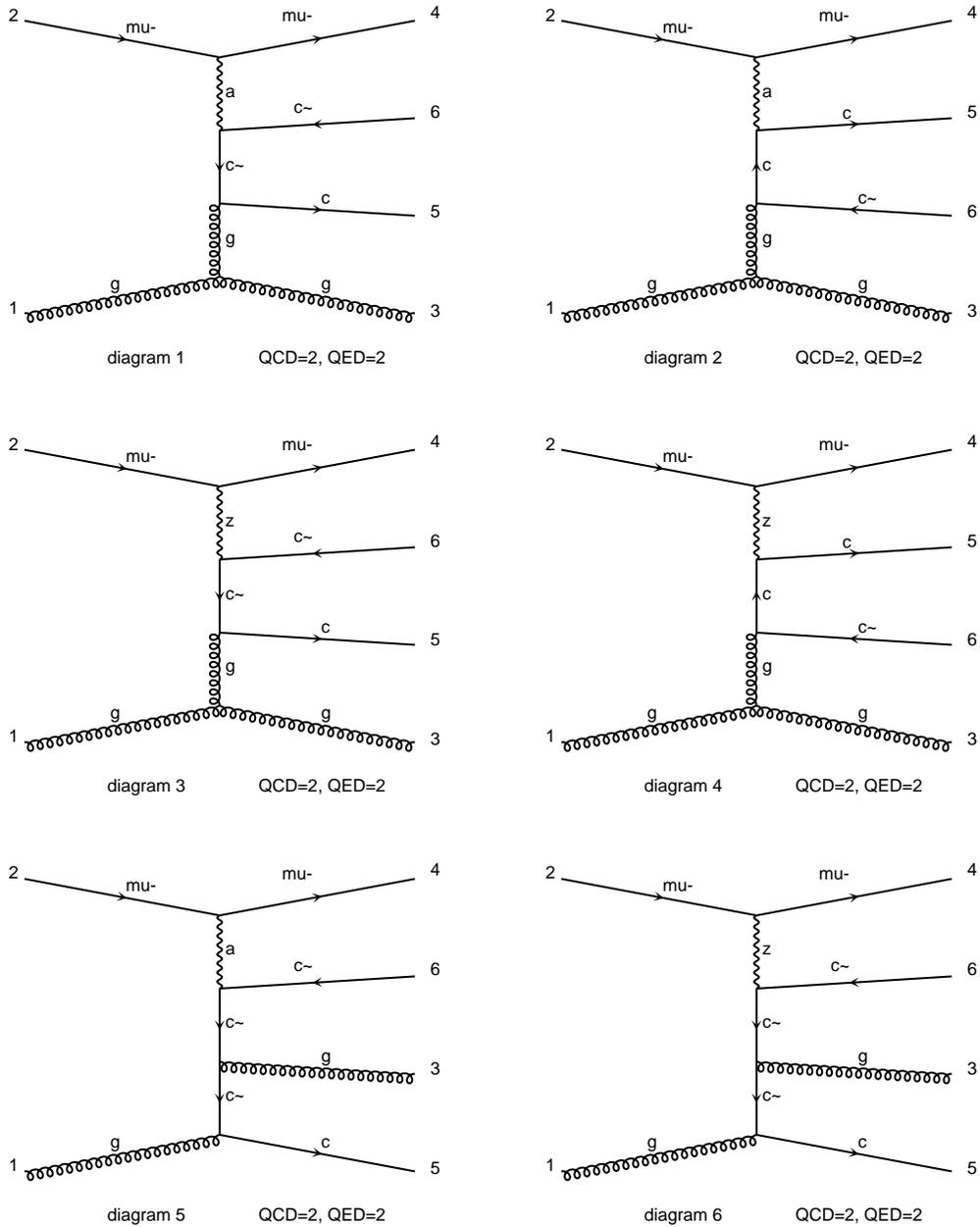}
    
    \caption{Diagrams used in the calculation of the NC DIS di-charm (non-resonant) process (SM5F\_NLO). The processes emphasized here are induced by gluons in the proton.}
    \label{fig:sm5f_nlo_dicharm_ncdis7}
\end{figure}

\clearpage

\subsection{Charged-Current DIS Di-Charm Production (No Higgs Resonance)}
\label{sec:ccdis_dicharm_mg5}

\begin{figure}[htbp]
    \centering
    \includegraphics[width=0.90\linewidth]{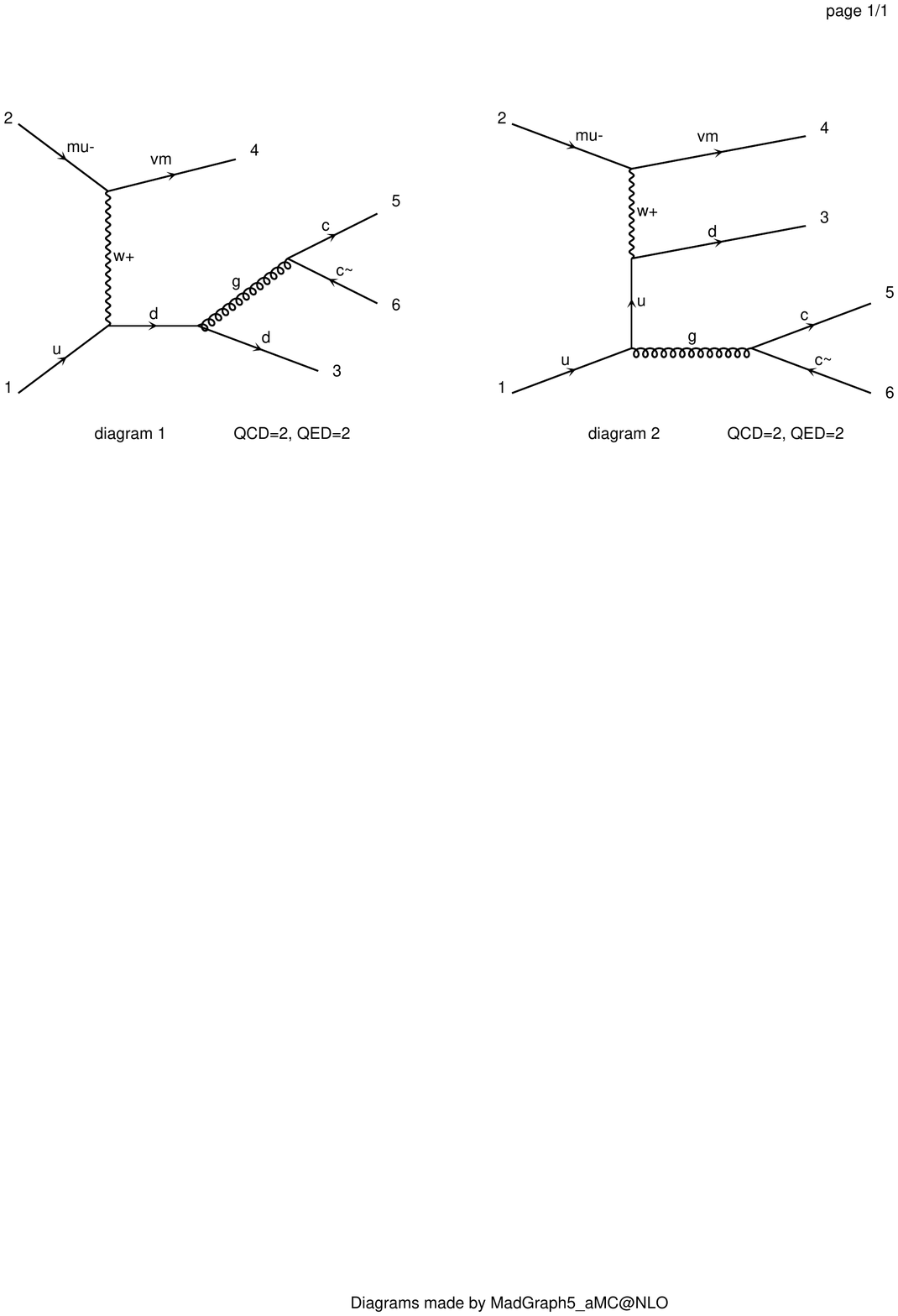}
    \caption{Diagrams used in the calculation of the CC DIS di-charm (non-resonant) process (SM5F\_NLO). The processes emphasized here are induced by up quarks in the proton.}
    \label{fig:sm5f_nlo_dicharm_ccdis1}
\end{figure}

\begin{figure}[htbp]    
    \includegraphics[width=0.90\linewidth]{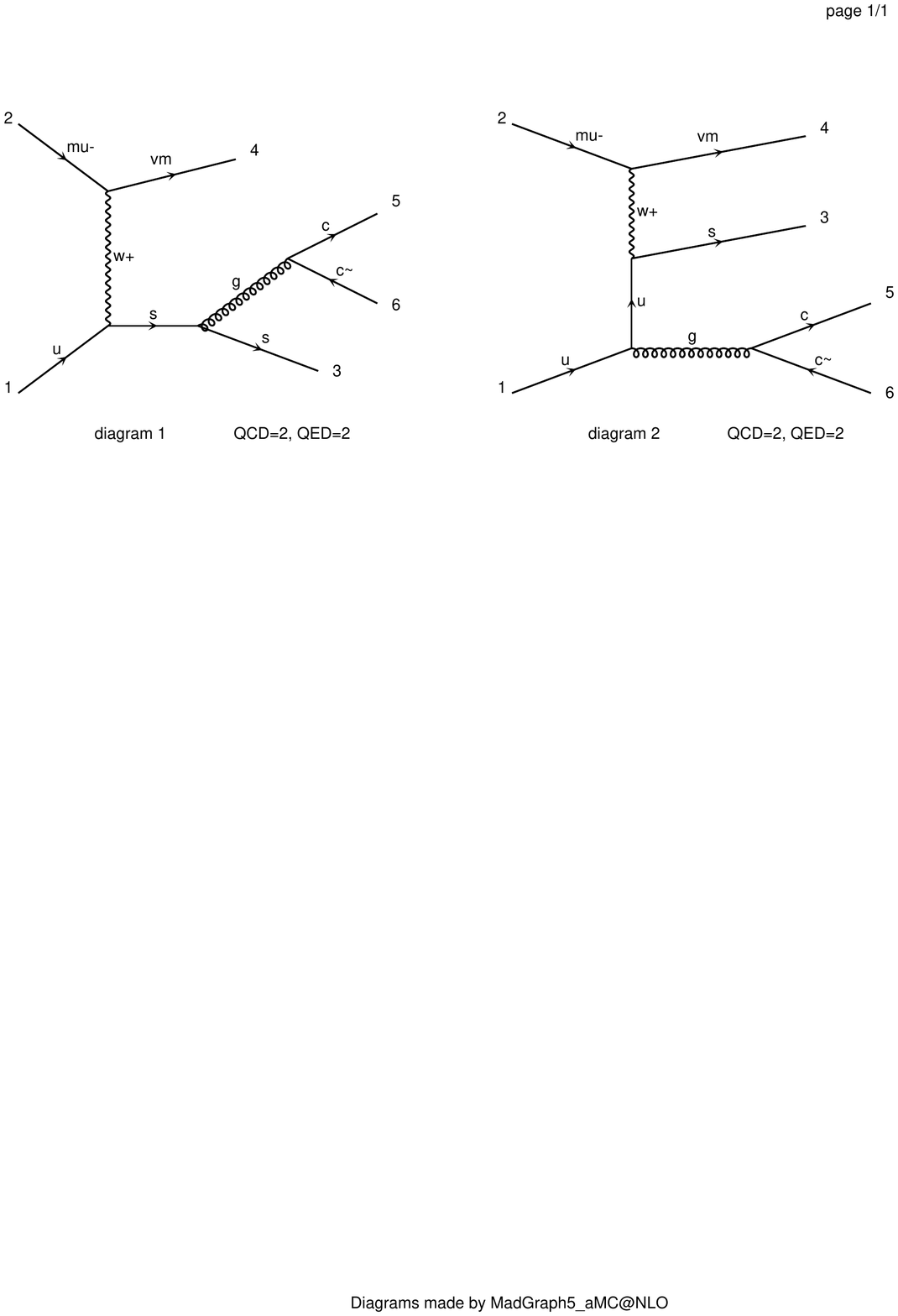}
    \caption{Diagrams used in the calculation of the CC DIS di-charm (non-resonant) process (SM5F\_NLO). The processes emphasized here are induced by up quarks in the proton.}
    \label{fig:sm5f_nlo_dicharm_ccdis2}
\end{figure}

\begin{figure}[htbp]    
    \includegraphics[width=0.90\linewidth]{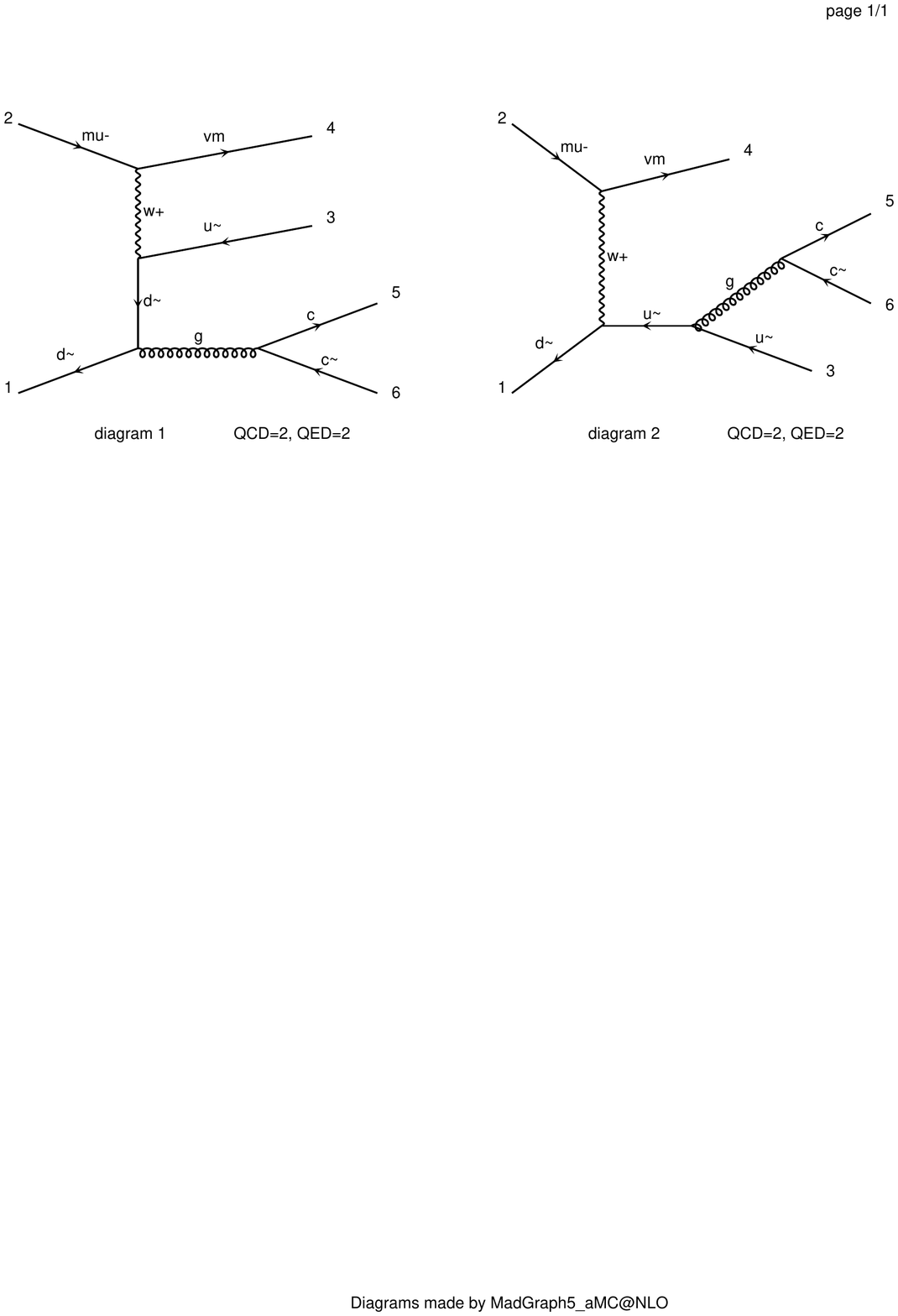}
    \caption{Diagrams used in the calculation of the CC DIS di-charm (non-resonant) process (SM5F\_NLO). The processes emphasized here are induced by down quarks in the proton.}
    \label{fig:sm5f_nlo_dicharm_ccdis3}
\end{figure}

\begin{figure}[htbp]    
    \includegraphics[width=0.90\linewidth]{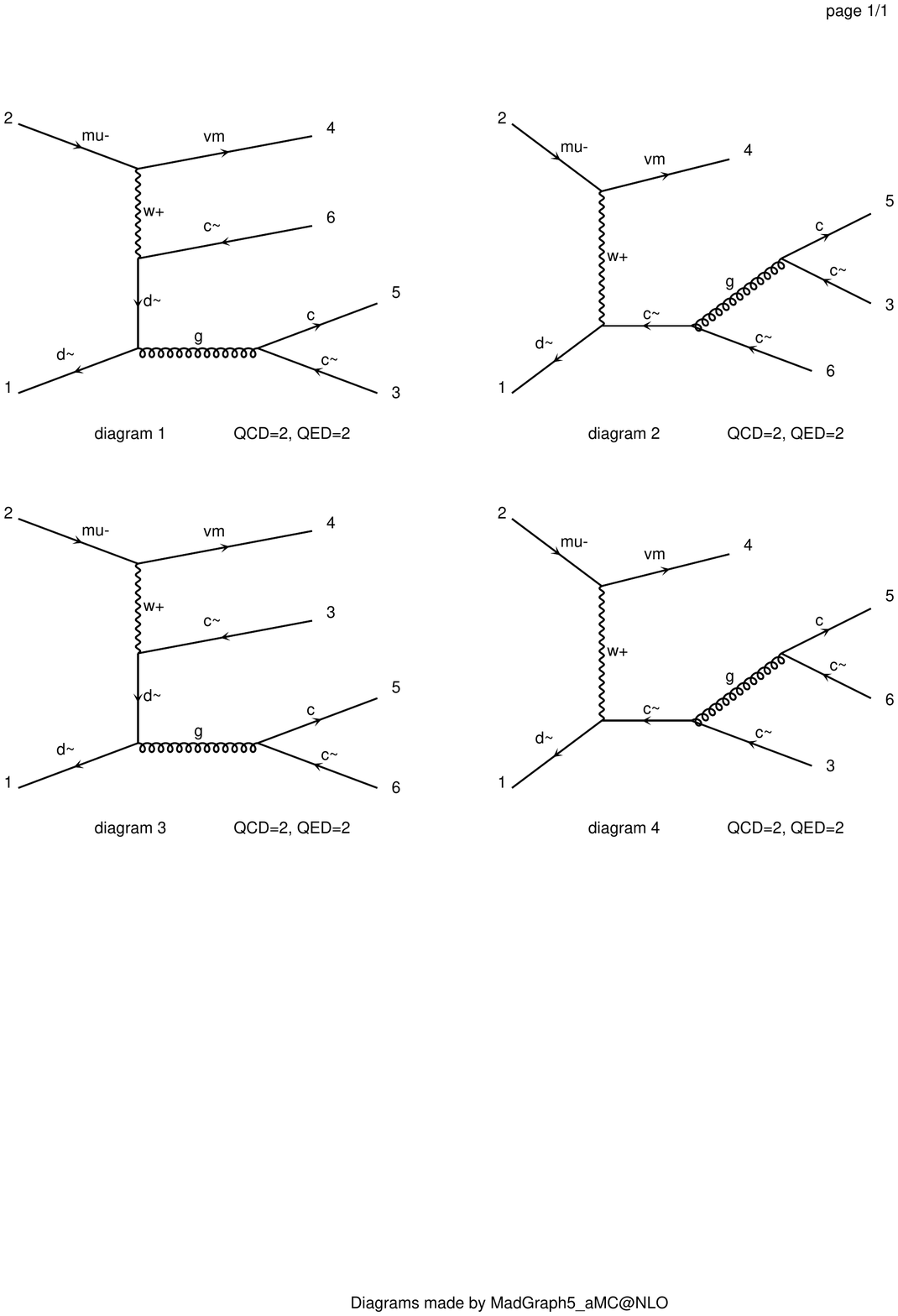}
    \caption{Diagrams used in the calculation of the CC DIS di-charm (non-resonant) process (SM5F\_NLO). The processes emphasized here are induced by down quarks in the proton.}
    \label{fig:sm5f_nlo_dicharm_ccdis4}
\end{figure}

\begin{figure}[htbp]
    \centering
    \includegraphics[width=0.90\linewidth]{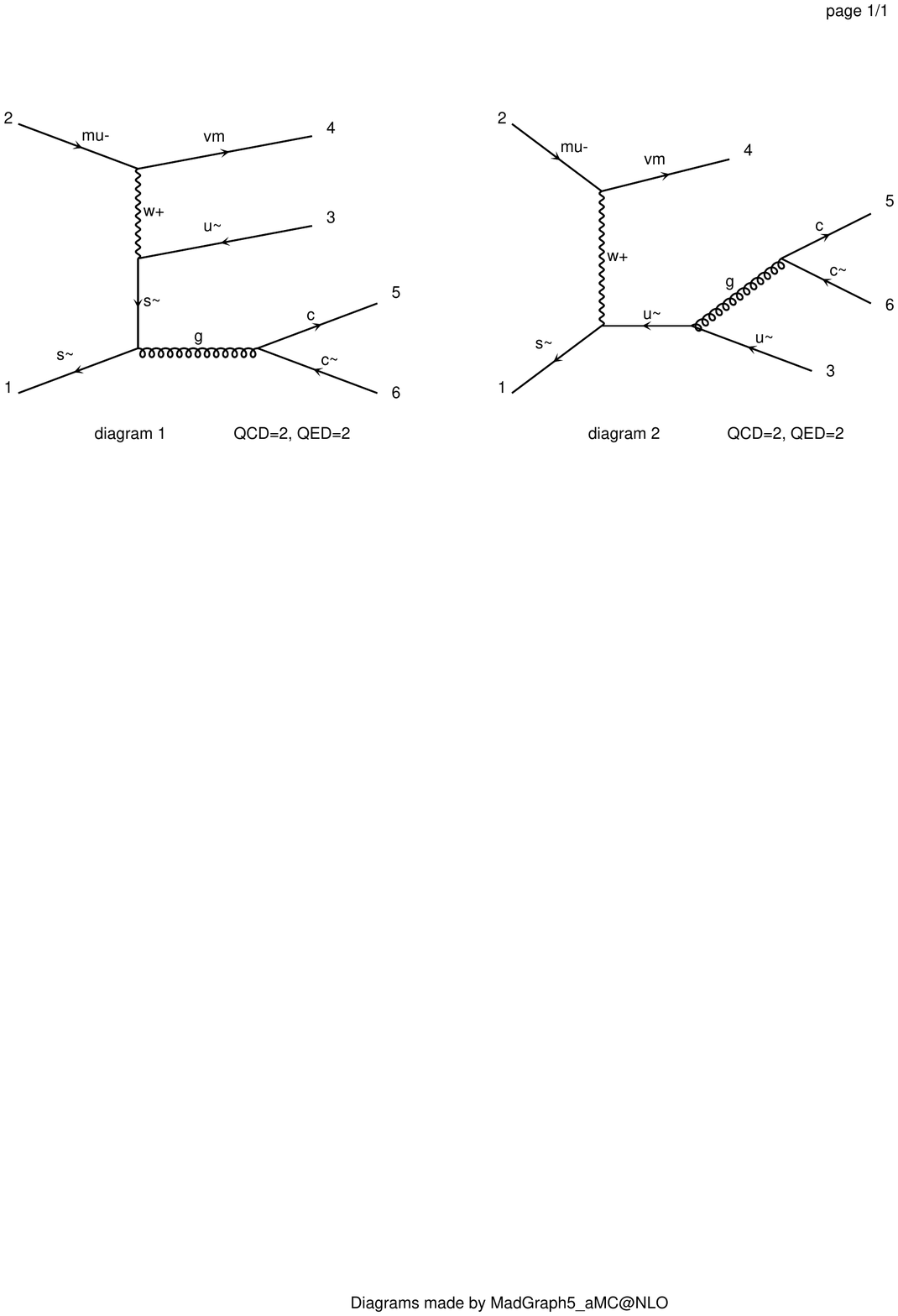}
    \caption{Diagrams used in the calculation of the CC DIS di-charm (non-resonant) process (SM5F\_NLO). The processes emphasized here are induced by strange quarks in the proton.}
    \label{fig:sm5f_nlo_dicharm_ccdis5}
\end{figure}

\begin{figure}[htbp]    
    \includegraphics[width=0.90\linewidth]{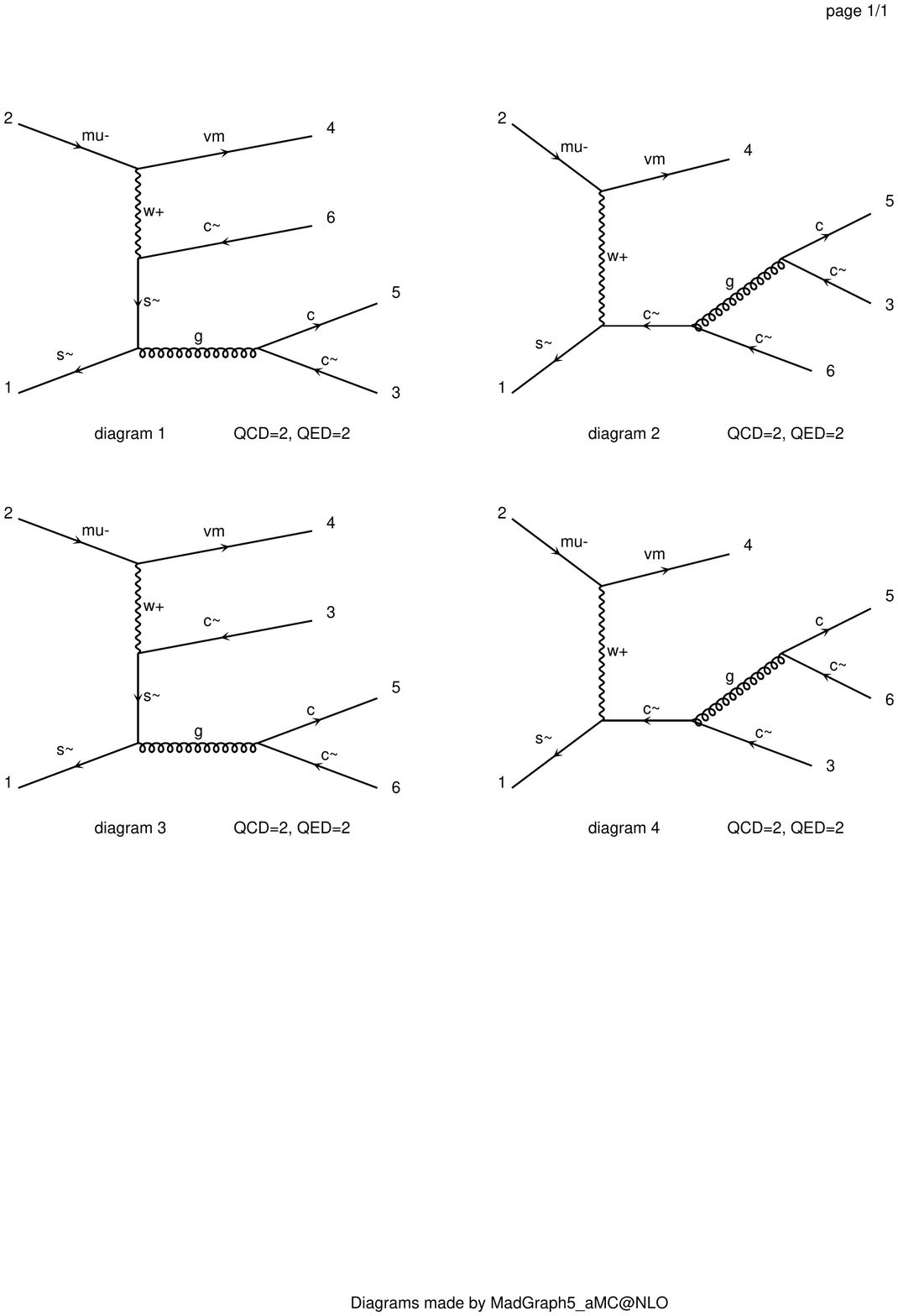}
    \caption{Diagrams used in the calculation of the CC DIS di-charm (non-resonant) process (SM5F\_NLO). The processes emphasized here are induced by strange quarks in the proton.}
    \label{fig:sm5f_nlo_dicharm_ccdis6}
\end{figure}

\begin{figure}[htbp]    
    \includegraphics[width=0.90\linewidth]{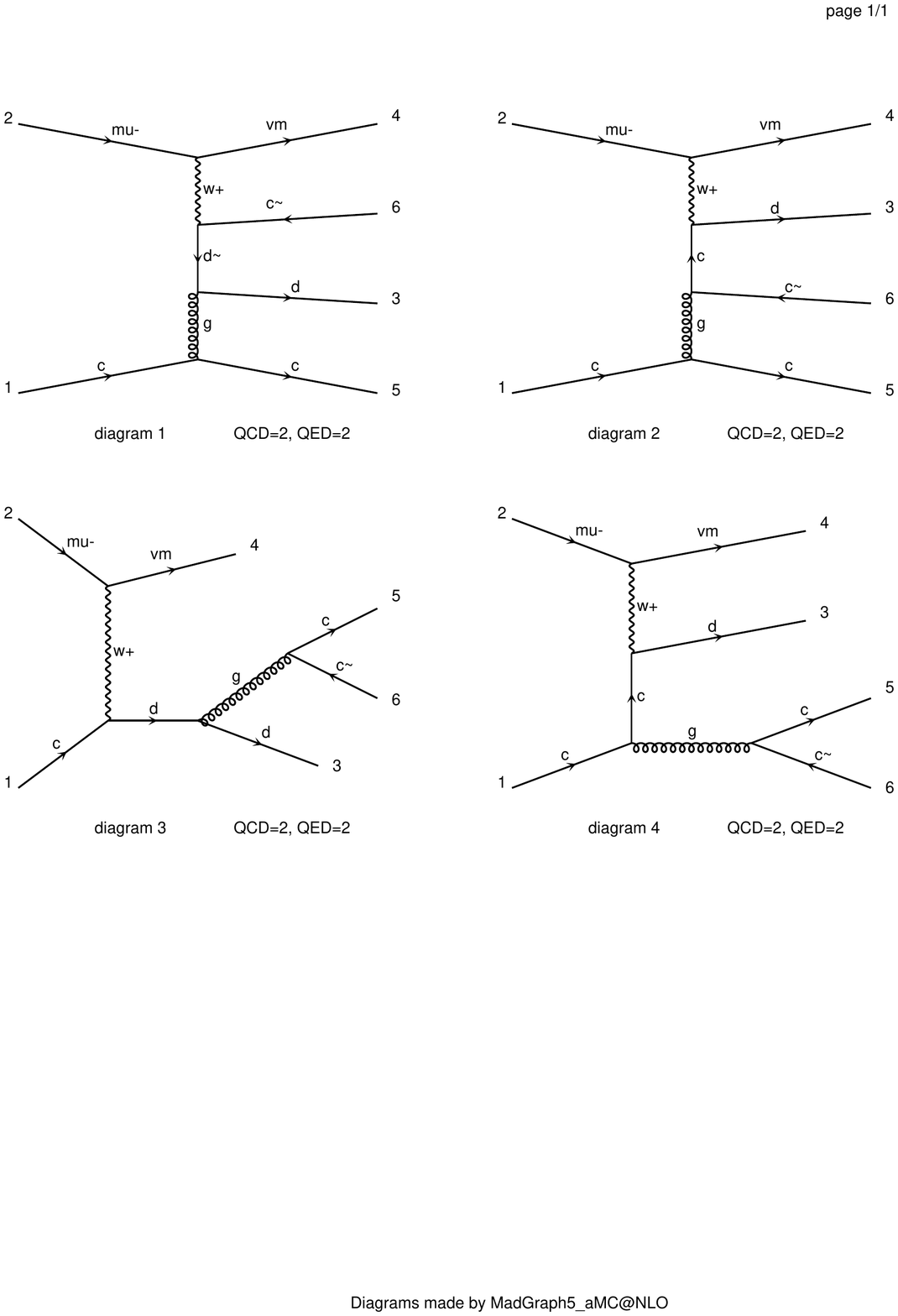}
    
    \caption{Diagrams used in the calculation of the CC DIS di-charm (non-resonant) process (SM5F\_NLO). The processes emphasized here are induced by charm quarks in the proton.}
    \label{fig:sm5f_nlo_dicharm_ccdis7}
\end{figure}

\begin{figure}[htbp]    
    \includegraphics[width=0.90\linewidth]{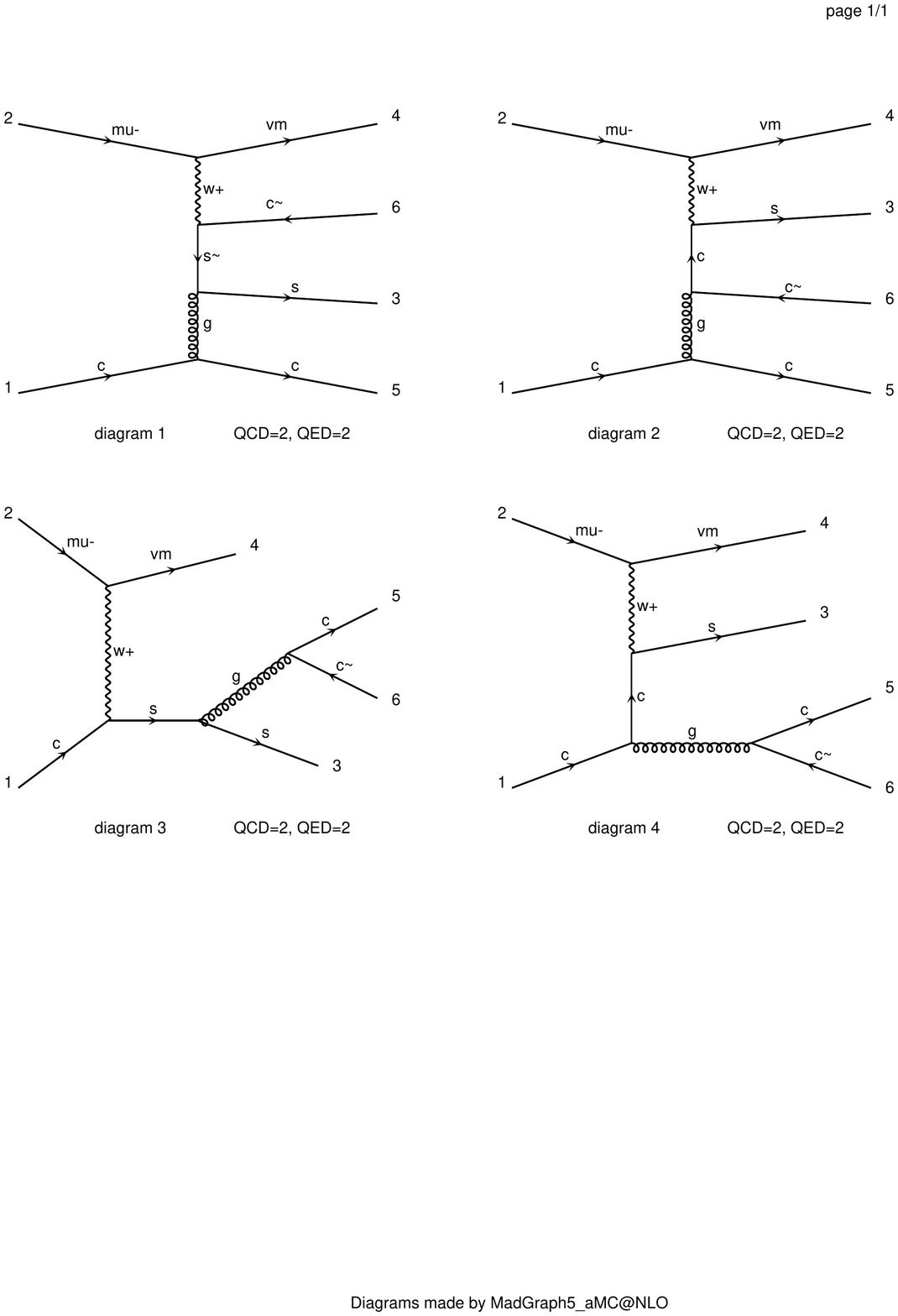}
    
    \caption{Diagrams used in the calculation of the CC DIS di-charm (non-resonant) process (SM5F\_NLO). The processes emphasized here are induced by charm quarks in the proton.}
    \label{fig:sm5f_nlo_dicharm_ccdis8}
\end{figure}

\twocolumngrid

\clearpage

\section{\mgamcnlo\ and \pythia\ Configurations}
\label{sec:simulation_details}

We here document, for the benefit of others who may wish to conduct similar simulations, the configuration of the matrix-element-level and hadronization-level simulation tools. All of these are achieved through the use of the \mgamcnlo\ interface, and the following code examples are provided assuming all simulation is run through this interface.

\subsection{Generator-level configuration}

The following code block illustrates the configuration of a proton-muon collider with the goal of producing the Higgs particle through CC DIS.

\begin{lstlisting}
set group_subprocesses False
define p = p b b~
define j = j b b~
import model SM5F_NLO
generate p mu- > j vm h
output pmu_275x960_ccdis_vbf_h-SM5F_NLO
\end{lstlisting}

The order of the initial- and final-state particles is important to the user-defined $Q^2$ calculation, defined below, as the calculation expects the initial lepton and the final lepton to have fixed positions in the generator string (e.g. the second and fourth positions).

The above configuration separates the subprocesses that result in the same final state; this is not observed to be necessary during signal simulation, but essential to the stability of the background simulation, which involves many more processes. We follow the procedure of Ref.~\cite{Acosta:2022ejc} and add bottom quark flavor to the proton definition, allowing it also to be produced in any explicit jet in the final state (e.g. through a flavor-changing or flavor-preserving process). This allows bottom quarks to be controlled by the PDF.

The follow generate strings illustrate how we produced the Higgs also through NC DIS, top-Higgs production, and how we produced di-charm background.

\begin{lstlisting}
# NC DIS Higgs Production
generate p mu- > j mu- h

# CC DIS di-charm Production, no Higgs
generate p mu- > j vm c c~ /h

# NC DIS di-charm Production, no Higgs
generate p mu- > j mu- c c~ /h
\end{lstlisting}

\subsection{Production-level configuration}

The following code block illustrates the production-level configuration that would be executed to make a larger number of simulated events, having defined the generator-level process. We employ the \texttt{CT18NNLO} PDF, which is \textsc{LHAPDF} ID 14000.

\begin{lstlisting}
launch pmu_275x960_ccdis_vbf_h-SM5F_NLO
shower=Pythia8
0
set nevents 20000
set ebeam1 275
set ebeam2 960
set polbeam1 0
set polbeam2 0
set pdlabel1 lhapdf
set lhaid 14000
set ptj 5
set ptl 0
set ptjmax -1
set ptlmax -1
set pt_min_pdg {}
set pt_max_pdg {}
set etaj -1
set etal -1
set etalmin 0
set eta_min_pdg {}
set eta_max_pdg {}
set mxx_min_pdg {}
set drjl 0
set drjlmax -1
set ptheavy 0
set maxjetflavor 5
set dynamical_scale_choice 0
#set systematics_program none
\end{lstlisting}

The dynamical scale choice enforces the user-level definition we provide below. The systematic error computation can be disabled if all one needs are generated events (e.g. if one is not interested in a production cycle in knowing the theory uncertainties on the cross-section).

\subsection{$Q^2$ definition}

We modify the file in the output folder for each process named \texttt{SubProcesses/setscales.f} and define the following user-defined scale:

\begin{lstlisting}
rscale = max(0d0,-sumdot(p(0,2),p(0,6),-1d0))
\end{lstlisting}
This assumes the the intial and final-state particle are the second and the 

\subsection{QCD color flow model in \pythia}

We also modify the color flow model for the beam remnant since the lepton-hadron collisions have been observed the pose a challenge to the default model. Specifically, we modify the \mgamcnlo\ \texttt{pythia8\_card\_default.dat} for each simulation process and add the following lines at the end:

\begin{lstlisting}
BeamRemnants:remnantMode = 1
ColourReconnection:mode = 1
\end{lstlisting}

\end{document}